%% file: main.tex
\documentclass[sigconf,nonacm]{acmart}

\usepackage{booktabs}
\usepackage{graphicx}
\usepackage{microtype}
\usepackage{subcaption}
\usepackage{tabularx}
\usepackage{xspace}
\usepackage{placeins}
\usepackage{xurl}
\usepackage[inline]{enumitem}
\usepackage{tikz}
\usepackage{pifont}
\usetikzlibrary{arrows.meta,positioning,fit,calc,patterns.meta,shapes.geometric}
\usepackage{pgfplots}
\pgfplotsset{compat=1.17}
\graphicspath{{figures/}}
\definecolor{textDark}{RGB}{38,51,59}
\definecolor{lineDark}{RGB}{96,112,120}
\definecolor{recordGreen}{RGB}{232,240,235}
\definecolor{evidenceBlue}{RGB}{232,239,244}
\definecolor{analysisGold}{RGB}{246,240,226}
\definecolor{holdoutRed}{RGB}{245,232,228}
\definecolor{plotTeal}{RGB}{38,104,110}
\definecolor{plotRust}{RGB}{181,82,54}
\definecolor{plotInk}{RGB}{55,69,76}
\definecolor{plotGray}{RGB}{205,211,213}
\definecolor{plotGold}{RGB}{178,140,40}

\newcommand{\papertitle}{A Difference Without a Distinction}
\newcommand{\papersubtitle}{A Measurement Study of \texorpdfstring{\\}{}Release Trust Evidence in Package Registries}

\input{generated/dynamics_macros}
\input{generated/dynamics_attribution_units_macros}

\input{generated/dynamics_pnpm_macros}

\input{generated/dynamics_case_macros}
\input{generated/dynamics_exploratory}
\input{generated/dynamics_strata_macros}
\input{generated/dynamics_pipeline_macros}

\input{generated/dynamics_eco_macros}
\input{generated/dynamics_pos_macros}
\input{generated/dynamics_credit_macros}
\input{generated/dynamics_frame4_macros}
\input{generated/dynamics_census_macros}

\input{generated/dynamics_line_macros}
\input{generated/dynamics_runstart_macros}
\input{generated/dynamics_ignoreafter_macros}

\input{generated/dynamics_adoption_macros}

\input{generated/dynamics_pnpmcurrent_macros}

\input{generated/dynamics_sigstoresupp_macros}
\input{generated/dynamics_signer_macros}
\input{generated/dynamics_denominators_macros}
\input{generated/dynamics_declarations_macros}

\input{generated/dynamics_declreplay_macros}
\input{generated/dynamics_declconv_macros}
\input{generated/dynamics_censuswaves_macros}

\input{generated/dynamics_signerregistry_macros}

\input{generated/dynamics_pnpmmid_macros}
\input{generated/dynamics_ecoday2_macros}

\input{generated/dynamics_facts_macros}
\input{generated/dynamics_supplementary_macros}

\input{generated/dynamics_mostdownloaded_macros}

\setcopyright{none}
\makeatletter
\let\hyxmp@parse@acmart\relax
\makeatother

\begin{document}

\title[\papertitle]{\papertitle\texorpdfstring{: }{: }\papersubtitle}

\author{Igor Santos-Grueiro}
\affiliation{%
  \institution{International University of La Rioja}
  \country{}
}

\begin{abstract}
\input{sections/abstract}
\end{abstract}



\maketitle

\input{sections/introduction}
\input{sections/background}
\input{sections/study_design}
\input{sections/results}
\input{sections/case_studies}
\input{sections/discussion}
\input{sections/related_work}
\input{sections/conclusion}

\bibliographystyle{ACM-Reference-Format}
\bibliography{refs}

\appendix
\input{sections/appendix}

\end{document}

%% file: generated/dynamics_macros.tex
\newcommand{\DynFOneCoordinates}{145}

\newcommand{\DynFTwoCoordinates}{295}

\newcommand{\DynFThreeCoordinates}{375}

\newcommand{\DynValRefReleases}{50}
\newcommand{\DynValRefChecks}{400}
\newcommand{\DynValCertChecked}{100}

%% file: generated/dynamics_attribution_units_macros.tex
\newcommand{\DynResidualAlerts}{321}
\newcommand{\DynResidualPublisherOrKeyAlerts}{274}

%% file: generated/dynamics_pnpm_macros.tex
\newcommand{\DynPnpmVersion}{10.29.3}
\newcommand{\DynPnpmConcordanceCases}{301}
\newcommand{\DynPnpmConcordanceAgree}{301}
\newcommand{\DynPnpmEmulationComparedVersions}{67{,}831}
\newcommand{\DynPnpmInFrameStableBlocksTotal}{82}
\newcommand{\DynPnpmInFrameStableBlocksTopPackage}{66}
\newcommand{\DynPnpmSampleSize}{200}
\newcommand{\DynPnpmBlockerNoEvidence}{82}
\newcommand{\DynStagedRankSince}{11.5.0}
\newcommand{\DynPnpmTrustedPublisherNeedsProvenanceSince}{11.4.0}
\newcommand{\DynPnpmCooldownDefaultSince}{11.0.0}

\newcommand{\DynPnpmLockfileVerifySince}{11.1.3}

\newcommand{\DynPnpmTrustLockfileSince}{11.3.0}

\newcommand{\DynPnpmExcludePruneSince}{12.4.0}

\newcommand{\DynPnpmIgnoreAfterSince}{10.27.0}

%% file: generated/dynamics_case_macros.tex
\newcommand{\DynCaseSemverAdvisoryCve}{CVE-2022-25883}

\newcommand{\DynCaseUuidAdvisoryCve}{CVE-2026-41907}
\newcommand{\DynCaseSemverWeeklyDownloads}{1{,}005{,}347{,}492}

%% file: generated/dynamics_pipeline_macros.tex
\newcommand{\DynPipeFTwoPackages}{295}
\newcommand{\DynPipeFTwoInterleavedPackages}{62}

\newcommand{\DynPipeFTwoStableBackportPackages}{29}

%% file: generated/dynamics_eco_macros.tex
\newcommand{\DynEcoPnpmVersion}{10.29.3}
\newcommand{\DynEcoTopTargets}{1{,}000}
\newcommand{\DynEcoTopFailing}{16}
\newcommand{\DynEcoTopFailShare}{1.6\%}
\newcommand{\DynEcoTopFailShareLo}{0.0\%}
\newcommand{\DynEcoTopFailShareHi}{4.5\%}
\newcommand{\DynEcoTopTransitiveBlocks}{15}
\newcommand{\DynEcoTopControlConfirms}{16}
\newcommand{\DynEcoRandomTargets}{997}
\newcommand{\DynEcoRandomFailing}{76}
\newcommand{\DynEcoRandomFailShare}{7.6\%}
\newcommand{\DynEcoRandomFailShareLo}{4.1\%}
\newcommand{\DynEcoRandomFailShareHi}{11.7\%}
\newcommand{\DynEcoRandomZeroDownloadShare}{28.4\%}
\newcommand{\DynEcoRandomTransitiveBlocks}{68}
\newcommand{\DynEcoRandomUnrelatedFailures}{15}
\newcommand{\DynEcoRandomControlConfirmed}{75}
\newcommand{\DynEcoDistinctBlockers}{34}
\newcommand{\DynEcoTotalFailing}{92}
\newcommand{\DynEcoTotalTargets}{1{,}997}
\newcommand{\DynEcoTopTopBlockerFailing}{14}
\newcommand{\DynEcoSecurityBackportReleases}{3}
\newcommand{\DynEcoSecurityBackportOneTargets}{29}

\newcommand{\DynEcoOlderLineBackportReleases}{10}
\newcommand{\DynEcoTopOlderLineBackportFailing}{14}
\newcommand{\DynEcoRandomOlderLineBackportFailing}{39}
\newcommand{\DynEcoBlockerNoEvidenceReleases}{31}

\newcommand{\DynEcoBlockerProvenanceAfterTpReleases}{3}

\newcommand{\DynEcoRandomPoolListed}{10{,}000}

\newcommand{\DynEcoRandomPoolWindowHours}{20.9}
\newcommand{\DynEcoRandomScopedShare}{75.1\%}
\newcommand{\DynEcoRandomScopes}{491}

\newcommand{\DynEcoRandomScopeUnitsFailShare}{8.3\%}

\newcommand{\DynEcoTopFailingRoleBuild}{4}
\newcommand{\DynEcoTopFailingRoleLint}{4}
\newcommand{\DynEcoTopFailingRoleTest}{8}

\newcommand{\DynEcoTopFailingRoleTooling}{16}

\newcommand{\DynEcoTopFailingReact}{4}

%% file: generated/dynamics_pos_macros.tex
\newcommand{\DynPosIncidents}{12}
\newcommand{\DynPosIncidentsNpm}{10}
\newcommand{\DynPosIncidentsPypi}{2}
\newcommand{\DynPosPackages}{15}
\newcommand{\DynPosMalReleases}{41}
\newcommand{\DynPosMalPublisherRecovered}{21}
\newcommand{\DynPosWindowDays}{90}
\newcommand{\DynPosROneAdjacentCaught}{5}

\newcommand{\DynPosROneHistoryCaught}{17}

\newcommand{\DynPosROneHistoryBenign}{60}

\newcommand{\DynPosROneHistoryLineCaught}{14}

\newcommand{\DynPosDeployedCaughtIncidents}{5}
\newcommand{\DynPosDeployedCaughtInferredEvidence}{2}

\newcommand{\DynPosMalPublisherInferred}{10}

\newcommand{\DynPosBenignFlagLitellmPypi}{56}
\newcommand{\DynPosAccessAccountIncidents}{6}

\newcommand{\DynPosAccessTokenIncidents}{6}

\newcommand{\DynPosBypassProvenanceReleases}{13}

\newcommand{\DynPosBypassTrustedPublisherReleases}{2}

\newcommand{\DynPosBypassWithAccessReleases}{15}
\newcommand{\DynPosBypassFlaggedReleases}{17}
\newcommand{\DynPubEmailHistoryAdded}{17}
\newcommand{\DynPubEmailHistoryAddedPackages}{12}

%% file: generated/dynamics_credit_macros.tex
\newcommand{\DynRegROneStableBlocksPypiFTwo}{6}
\newcommand{\DynRegROneStableBlocksCratesFTwo}{3}
\newcommand{\DynRegROneStableBlocksPypiFThree}{66}

%% file: generated/dynamics_frame4_macros.tex
\newcommand{\DynFFourPipePackages}{366}

\newcommand{\DynFFourCollectionDateProse}{September 27, 2026}

%% file: generated/dynamics_census_macros.tex
\newcommand{\DynCensusMinPrior}{3}
\newcommand{\DynCensusPriorGapDays}{30}
\newcommand{\DynCensusAllPackages}{1{,}654}
\newcommand{\DynCensusAllPriorEvidence}{289}
\newcommand{\DynCensusAllPriorEvidenceShare}{17.5\%}

\newcommand{\DynCensusAllMalVersions}{4{,}229}
\newcommand{\DynCensusAllMalFlagged}{507}
\newcommand{\DynCensusAllMalFlaggedShare}{12.0\%}

\newcommand{\DynCensusAllIncidents}{33}
\newcommand{\DynCensusAllIncidentsPriorEvidence}{17}

%% file: generated/dynamics_line_macros.tex
\newcommand{\DynLineBackportsTotal}{3}
\newcommand{\DynLineBackportsPassPerLineVacuous}{2}

%% file: generated/dynamics_ignoreafter_macros.tex
\newcommand{\DynIgnoreAfterPnpmVersion}{10.29.3}
\newcommand{\DynIgnoreAfterRandomFailingNone}{77}

\newcommand{\DynIgnoreAfterRandomFailingThirtyDays}{19}

\newcommand{\DynIgnoreAfterPosReleases}{41}

\newcommand{\DynIgnoreAfterPosFlagged}{17}
\newcommand{\DynIgnoreAfterPosMedianLiveHours}{3.1}
\newcommand{\DynIgnoreAfterPosMaxLiveDays}{2.5}
\newcommand{\DynIgnoreAfterTopMinBlockerAgeDays}{629.2}

\newcommand{\DynIgnoreAfterCooldownKeptOutOneDay}{36}

\newcommand{\DynIgnoreAfterCooldownFlaggedOutliveOneDay}{5}
\newcommand{\DynIgnoreAfterCooldownOutliveOneDay}{5}
\newcommand{\DynIgnoreAfterCooldownOutliveUnknownRemovalOneDay}{1}

\newcommand{\DynIgnoreAfterCooldownFlaggedKeptOutOneDay}{12}

\newcommand{\DynIgnoreAfterCooldownKeptOutSevenDays}{40}

%% file: generated/dynamics_adoption_macros.tex
\newcommand{\DynAdoptionEnabling}{4{,}516}

\newcommand{\DynAdoptionHeadlineRepos}{4{,}516}

\newcommand{\DynAdoptionHeadlineWithExcludeShare}{37.1\%}

\newcommand{\DynAdoptionHeadlineReposWhole}{290}

\newcommand{\DynAdoptionTopOnePackages}{1{,}172}

\newcommand{\DynAdoptionSemverSixThreeOne}{994}
\newcommand{\DynAdoptionCooldownWithIgnoreAfter}{588}
\newcommand{\DynAdoptionIgnoreAfterLeCooldown}{157}

\newcommand{\DynAdoptionOwners}{2{,}645}

%% file: generated/dynamics_pnpmcurrent_macros.tex
\newcommand{\DynPnpmCurrentVersion}{12.9.1}
\newcommand{\DynPnpmCurrentLegacyVersion}{10.29.3}
\newcommand{\DynPnpmCurrentObservedDateProse}{October 4, 2026}
\newcommand{\DynPnpmCurrentDefaultTopFailing}{15}

\newcommand{\DynPnpmCurrentDefaultTopFallback}{1}

\newcommand{\DynPnpmCurrentDefaultRandomFailing}{58}

\newcommand{\DynPnpmCurrentDefaultRandomFallback}{12}

\newcommand{\DynPnpmCurrentLegacyRandomPassingNowFailing}{7}
\newcommand{\DynPnpmCurrentDefaultTopFailingOnTopBlocker}{14}

\newcommand{\DynPnpmCurrentNowFailingCauseTpNoProvenance}{2}

\newcommand{\DynPnpmCurrentNowFailingCauseStaged}{5}

\newcommand{\DynPnpmCurrentNowFailingStagedWithTrustedPublisher}{3}

\newcommand{\DynPnpmCurrentNowFailingStagedWithoutEvidence}{2}

%% file: generated/dynamics_sigstoresupp_macros.tex
\newcommand{\DynSupplCollectedOnProse}{October 4, 2026}

%% file: generated/dynamics_signer_macros.tex
\newcommand{\DynSignerFOneSourceDiffReleases}{41}
\newcommand{\DynSignerFOneSourceDiffPackages}{1}
\newcommand{\DynSignerFTwoTagWildcardTimestampAdjacentAlerts}{232}

\newcommand{\DynSignerFTwoPartAlerts}{690}

\newcommand{\DynSignerPooledTwoKeylessPackages}{165}
\newcommand{\DynSignerPooledAdjacentFullSanPackages}{125}
\newcommand{\DynSignerPooledAdjacentFullSanShare}{75.8\%}

\newcommand{\DynSignerPooledAdjacentSanWithoutRefPackages}{23}
\newcommand{\DynSignerPooledAdjacentSanWithoutRefShare}{13.9\%}

%% file: generated/dynamics_denominators_macros.tex
\newcommand{\DynProvLossPooledLost}{19}
\newcommand{\DynProvLossPooledEver}{188}
\newcommand{\DynProvLossPooledShare}{10.1\%}
\newcommand{\DynProvLossPooledShareLow}{6.6\%}
\newcommand{\DynProvLossPooledShareHigh}{15.2\%}
\newcommand{\DynFramesFTwoNpmPopularPool}{300}

\newcommand{\DynFramesFTwoNpmRandomPool}{4{,}000}

\newcommand{\DynFramesFTwoPopularPassRate}{50.2\%}
\newcommand{\DynFramesFTwoRandomPassRate}{13.7\%}
\newcommand{\DynFramesPopularExcludedMinFTwo}{20}
\newcommand{\DynFramesPopularExcludedMaxFTwo}{187}
\newcommand{\DynFramesPooledNpmPopularKeyless}{51.4\%}

\newcommand{\DynFramesPooledNpmRandomKeyless}{25.9\%}

\newcommand{\DynFramesPooledNpmKeylessDiff}{25.5}
\newcommand{\DynFramesPooledNpmKeylessDiffLow}{12.7}
\newcommand{\DynFramesPooledNpmKeylessDiffHigh}{37.1}

\newcommand{\DynDecompFOneRTwoRefMinShare}{56.4\%}

%% file: generated/dynamics_declarations_macros.tex
\newcommand{\DynDeclKeysPooledApacheAlerts}{30}

\newcommand{\DynDeclKeysPooledDeclaredAlerts}{29}

\newcommand{\DynDeclKeysPooledOtherAlerts}{27}

\newcommand{\DynDeclCratesAllWithTrustedPublishing}{26}
\newcommand{\DynDeclCratesAllTrustpubOnlyOfTrustedPublishing}{18}

%% file: generated/dynamics_declreplay_macros.tex
\newcommand{\DynDeclReplayEcoReleases}{34}
\newcommand{\DynDeclReplayIncidentWithEarlierEvidence}{17}
\newcommand{\DynDeclReplayIncidentAccountReleases}{21}

%% file: generated/dynamics_declconv_macros.tex
\newcommand{\DynDeclReplayAsReplayedEcoLineHistoryFailuresRemovedBoth}{59}

\newcommand{\DynDeclReplayAsReplayedEcoRegisteredRefuse}{31}
\newcommand{\DynDeclReplayAsReplayedEcoPerLineFailuresRemovedBoth}{53}

\newcommand{\DynDeclReplayAsReplayedIncidentPerLineAdmitWithEarlierEvidence}{3}

\newcommand{\DynDeclReplayEnforcingEcoRegisteredRefuse}{33}
\newcommand{\DynDeclReplayEnforcingEcoPerLineFailuresRemovedBoth}{47}

\newcommand{\DynDeclReplayEnforcingIncidentPerLineAdmitWithEarlierEvidence}{2}

\newcommand{\DynDeclReplayEcoFailuresBoth}{92}
\newcommand{\DynDeclReplayIncidentWithoutEarlierEvidence}{24}
\newcommand{\DynDeclReplayIncidentWithoutEarlierEvidencePackages}{10}
\newcommand{\DynDeclReplayIncidentWithoutEarlierEvidenceBeforeNpmProvenance}{13}
\newcommand{\DynDeclReplayIncidentNoTokenRefuse}{20}
\newcommand{\DynDeclReplayEnforcingRepositoryDecided}{2}

\newcommand{\DynDeclReplayWindowBenign}{874}
\newcommand{\DynDeclReplayWindowBenignNoEvidence}{367}

\newcommand{\DynDeclReplaySampleStableNpmReleases}{18{,}225}

\newcommand{\DynDeclReplaySampleStableNpmPackages}{249}
\newcommand{\DynDeclReplaySampleStableNpmPackagesNoPipeline}{145}

\newcommand{\DynDeclReplayAsReplayedSampleStableNpmDeployedRefuse}{83}

\newcommand{\DynDeclReplayAsReplayedSampleStableNpmRegisteredRefuse}{16{,}235}

\newcommand{\DynDeclReplayAsReplayedSampleStableNpmPerLineRefuse}{43}

\newcommand{\DynDeclReplayAsReplayedSampleStableNpmPerLineAdmitDeployedRefuses}{48}

\newcommand{\DynDeclReplayEnforcingSampleStableNpmRegisteredRefuse}{152}
\newcommand{\DynDeclReplayEnforcingSampleStableNpmRegisteredRefusePassesDeployed}{69}

\newcommand{\DynDeclReplayEnforcingSampleStableNpmPerLineRefuse}{143}

\newcommand{\DynDeclReplayEnforcingSampleStableNpmPerLineAdmitDeployedRefuses}{1}

\newcommand{\DynDeclReplaySampleStablePypiReleases}{10{,}013}

\newcommand{\DynDeclReplayAsReplayedSampleStablePypiDeployedRefuse}{98}

\newcommand{\DynDeclReplayAsReplayedSampleStablePypiRegisteredRefuse}{6{,}904}

\newcommand{\DynDeclReplayAsReplayedSampleStablePypiPerLineRefuse}{76}

\newcommand{\DynDeclReplayEnforcingSampleStablePypiRegisteredRefuse}{255}

\newcommand{\DynDeclReplayEnforcingSampleStablePypiPerLineRefuse}{168}

%% file: generated/dynamics_censuswaves_macros.tex
\newcommand{\DynCensusByWaveWaves}{17}

\newcommand{\DynCensusByWaveLargestCount}{5}
\newcommand{\DynCensusByWaveLargestShareMin}{0.8\%}
\newcommand{\DynCensusByWaveLargestShareMax}{97.7\%}
\newcommand{\DynCensusTopTwoPriorEvidenceShare}{51.9\%}

\newcommand{\DynCensusTopTwoMalFlaggedShare}{69.8\%}
\newcommand{\DynCensusWithoutTopTwoPriorEvidenceShare}{9.3\%}
\newcommand{\DynCensusIncidentBootPriorEvidenceLo}{6.5\%}
\newcommand{\DynCensusIncidentBootPriorEvidenceHi}{43.9\%}
\newcommand{\DynCensusMalEvidenceVersions}{119}
\newcommand{\DynCensusMalEvidencePackages}{46}

%% file: generated/dynamics_signerregistry_macros.tex
\newcommand{\DynSignerRegNpmPooledTagOnlyShare}{17.7\%}

\newcommand{\DynSignerRegPypiPooledTagOnlyShare}{74.4\%}

\newcommand{\DynSignerBranchPooledPackages}{23}
\newcommand{\DynSignerBranchPooledLine}{15}

%% file: generated/dynamics_pnpmmid_macros.tex
\newcommand{\DynPnpmMidVersion}{12.6.0}

\newcommand{\DynPnpmMidTopFailing}{16}

\newcommand{\DynPnpmMidLegacyTopFailing}{16}

\newcommand{\DynPnpmMidRandomFailing}{77}

\newcommand{\DynPnpmMidLegacyRandomFailing}{76}

\newcommand{\DynPnpmMidLegacyFailingRandomNowPass}{5}
\newcommand{\DynPnpmMidLegacyFailingRandomNowOther}{1}
\newcommand{\DynPnpmMidLegacyPassingRandomNowFailing}{7}
\newcommand{\DynPnpmMidPinsExact}{3{,}441}

\newcommand{\DynPnpmMidPinsOriginalShare}{34.4\%}
\newcommand{\DynPnpmMidPinsEarlyElevenShare}{9.5\%}
\newcommand{\DynPnpmMidPinsLaterScaleShare}{54.0\%}
\newcommand{\DynPnpmMidPinsOlderVersionShare}{2.1\%}

%% file: generated/dynamics_ecoday2_macros.tex
\newcommand{\DynEcoDayTwoDateProse}{October 4, 2026}

\newcommand{\DynEcoDayTwoTargets}{1{,}000}

\newcommand{\DynEcoDayTwoFailShare}{9.2\%}
\newcommand{\DynEcoDayTwoFailShareLo}{5.4\%}
\newcommand{\DynEcoDayTwoFailShareHi}{13.6\%}

\newcommand{\DynEcoDayTwoFailingOnDayOneBlockersShare}{63.0\%}

%% file: generated/dynamics_facts_macros.tex
\newcommand{\DynEvidNpmTopThreeShareMin}{59.8\%}
\newcommand{\DynEvidNpmTopThreeShareMax}{70.9\%}
\newcommand{\DynPosNotReplayedHeadlineIncidents}{2}

\newcommand{\DynPosNotReplayedHeadlineMalVersions}{8}

\newcommand{\DynEthicsWorkersMax}{16}

%% file: generated/dynamics_supplementary_macros.tex
\newcommand{\DynSignerRepoRepositories}{150}

\newcommand{\DynSignerRepoAllFullSanShareLow}{68.5\%}
\newcommand{\DynSignerRepoAllFullSanShareHigh}{82.5\%}
\newcommand{\DynSignerRepoNpmFullSanShare}{67.1\%}

\newcommand{\DynSignerRepoPypiFullSanShare}{83.7\%}

\newcommand{\DynSignerRepoDiffFullSan}{$-$16.6}
\newcommand{\DynSignerRepoDiffFullSanLow}{$-$31.9}
\newcommand{\DynSignerRepoDiffFullSanHigh}{$-$2.4}

\newcommand{\DynSignerRepoNpmTagWildcardShare}{48.1\%}

\newcommand{\DynSignerRepoPypiTagWildcardShare}{14.0\%}

\newcommand{\DynSignerRepoDiffTagWildcard}{34.1}
\newcommand{\DynSignerRepoDiffTagWildcardLow}{16.7}
\newcommand{\DynSignerRepoDiffTagWildcardHigh}{49.6}

\newcommand{\DynCensusWaveMinPackages}{3}
\newcommand{\DynCensusWaveGapHours}{72}
\newcommand{\DynCensusWindowDayHours}{24}

\newcommand{\DynCensusWindowDayPriorEvidenceLo}{7.2\%}
\newcommand{\DynCensusWindowDayPriorEvidenceHi}{43.7\%}

\newcommand{\DynCensusWindowWeekPriorEvidenceLo}{3.7\%}
\newcommand{\DynCensusWindowWeekPriorEvidenceHi}{31.8\%}

\newcommand{\DynAdoptionWebFrameworkShare}{42.1\%}

\newcommand{\DynAdoptionWebFrameworkFrameworks}{17}
\newcommand{\DynPypiFilesAttestedReleases}{3{,}656}
\newcommand{\DynPypiFilesMixedReleases}{1}
\newcommand{\DynPypiFilesFileCap}{10}

\newcommand{\DynDeclCratesAfterAdoptionCrates}{26}
\newcommand{\DynDeclCratesAfterAdoptionReleases}{519}
\newcommand{\DynDeclCratesAfterAdoptionTokenReleases}{3}

%% file: generated/dynamics_mostdownloaded_macros.tex
\newcommand{\DynEcoMostDownloadedCovered}{1{,}000}

\newcommand{\DynEcoMostDownloadedVersion}{10.29.3}

\newcommand{\DynEcoMostDownloadedObservedDateProse}{October 11, 2026}

\newcommand{\DynEcoMostDownloadedIsLatest}{540}

\newcommand{\DynEcoMostDownloadedFailing}{27}

\newcommand{\DynEcoMostDownloadedFailingLatest}{15}

\newcommand{\DynEcoMostDownloadedControlConfirms}{27}

%% file: sections/abstract.tex
A stolen npm token or account lets an attacker publish a malicious
version of a popular package. Such uploads skip the project's CI
workflow and can lack its provenance, so since November 2025 an opt-in
pnpm policy refuses any release with weaker evidence than an earlier
one. Maintainers also publish backports by hand; how often the policy
blocks them and which attacks it flags is unknown. We test the policy
on the latest versions of \DynEcoTotalTargets{} popular and recently
updated npm packages, replay its variants and identity pins on
reconstructed stolen-credential incidents, and analyze an npm malware
advisory census. We also compare
publication evidence across five package registries: npm, PyPI, Maven
Central, crates.io, and RubyGems. With
pnpm~\DynEcoPnpmVersion{},
\DynEcoTopFailShare{} of popular and \DynEcoRandomFailShare{} of recently
updated installs are blocked, mostly by backports, and one security fix
backported by hand, \texttt{semver}~\texttt{6.3.1}, causes
\DynEcoTopTopBlockerFailing{} of the \DynEcoTopFailing{} popular failures.
The policy can flag an attack only if an earlier release of the same
package carried stronger evidence. In the census,
\DynCensusAllPriorEvidenceShare{} of attacked packages had earlier
evidence; two campaigns supply about half of these. Attackers with the
observed access could have passed
\DynPosBypassWithAccessReleases{} of the \DynPosBypassFlaggedReleases{}
flagged releases by adding evidence. Pins on the full signing identity
alert on \DynSignerPooledAdjacentFullSanShare{} of packages with at least two
keyless-signed releases, largely because certificates name the Git
branch or tag. No variant tells a recent backport and an attack apart:
comparing per line or declaring manual lines at the registry admits
both, and an age cutoff exempts only old releases.

%% file: sections/introduction.tex
\section{Introduction}\label{sec:introduction}

On March~31, 2026, an attacker published two malicious versions of
\texttt{axios}, a widely used JavaScript HTTP client, from its lead
maintainer's compromised npm account~\cite{axios2026postmortem}. Similar
attacks affected other npm and PyPI
packages~\cite{microsoft2025shaihulud,pypi2026litellm_telnyx}. Such
attacks can reach popular applications through their dependencies and the
tools used to build and test them~\cite{zimmermann2019}.

Package managers can check a release's publication records at its
package registry before installation. This \emph{publication evidence}
includes publishing accounts, continuous integration (CI) workflows,
attestations, and signing keys (\S\ref{sec:evidence}). A release uploaded
with stolen credentials can lack evidence left by the project's usual
CI workflow. In November 2025, the package manager pnpm introduced an
opt-in \texttt{trustPolicy: no-downgrade} setting that refuses a version with weaker
evidence than an earlier version of the same
package~\cite{pnpm_trust_policy}. We found \DynAdoptionEnabling{} public
GitHub repositories of \DynAdoptionOwners{} owners that enable it;
\DynAdoptionWebFrameworkShare{} of their root manifests declare a Web UI
framework.

The same check can block a security update. A pnpm user reported in a
GitHub issue that pnpm refused \texttt{semver} version~\texttt{6.3.1}, a fix
backported by hand to an older major line, and asked for a comparison
per major version~\cite{pnpm_issue_10202}. A pnpm maintainer
answered that a stolen token can publish on any line, so only the publish
date orders releases reliably. An age cutoff
proposed there, which exempts old releases, shipped in
pnpm~\DynPnpmIgnoreAfterSince{}~\cite{pnpm_pr_10359}. In our adoption
sample, \DynAdoptionSemverSixThreeOne{} repositories exempt
\texttt{semver}~\texttt{6.3.1}. A manual backport and a takeover can show
the same evidence drop. The records show how publication changed;
that change alone does not establish whether the maintainer authorized
the release.

Prior work detects malicious packages from registry
metadata~\cite{halder2024metadata} and studies whether, how well, and why
releases are
signed~\cite{schorlemmer2024registrysigning,kalu2025industrysigning,kalu2026usability}.
How often pnpm's policy blocks installs, which attacks it flags, and what
relaxing it gives up remain unanswered. We ask three questions about
\emph{consumer rules} that compare publication evidence across releases:
\begin{enumerate*}[label=\textbf{RQ\arabic*:}]
  \item What does pnpm's policy cost, and which attacks does it flag?
  \item Which design choices control these outcomes, and what would a
  declared policy change?
  \item What evidence do package registries expose, and how do identity pins
  behave?
\end{enumerate*}

For RQ1, we run three pnpm releases with the policy on the latest
versions of \DynEcoTopTargets{} popular and \DynEcoRandomTargets{}
recently updated npm packages (\S\ref{sec:ecosystem}) and replay its
rule offline on \DynPosIncidents{} reconstructed stolen-credential
incidents and a census of npm malware advisories
(\S\ref{sec:positive-controls}). For RQ2, we vary the history scope,
age cutoff, and evidence ranking, and replay two policies that a package could declare at its registry. For
RQ3, we compare the evidence available to these rules across npm,
PyPI, Maven Central, crates.io, and RubyGems. We collect release histories
from April 2024 to September 2026 for four samples and replay the policy
and three \emph{identity pins}, which alert on a
changed signer, key, or publisher (\S\ref{sec:samples}).

With pnpm~\DynEcoPnpmVersion{}, \DynEcoTopFailing{} of the
\DynEcoTopTargets{} most-downloaded npm packages fail to install at their
latest version, and \DynEcoMostDownloadedFailing{} at their
most-downloaded version (\S\ref{sec:rq1}). All \DynEcoTopFailing{} are
JavaScript build, lint, or test tools, including Babel plugins and ESLint
packages, and \DynEcoTopFailingReact{} target React. One security
backport, \texttt{semver}~\texttt{6.3.1}, causes \DynEcoTopTopBlockerFailing{} of
the \DynEcoTopFailing{} failures (\S\ref{sec:case-studies}). The policy
can flag an attack only where earlier releases carried evidence, as for
\DynCensusAllPriorEvidenceShare{} of attacked packages in our census
(\S\ref{sec:rq1}), and attackers could have passed
\DynPosBypassWithAccessReleases{} of the \DynPosBypassFlaggedReleases{}
malicious releases that pnpm flags by adding evidence. A history per
maintenance line admits backports and attacks alike (\S\ref{sec:rq2}):
it passes \texttt{axios} version~\texttt{0.30.4}, the attacker's release on an older line,
which names the same account as the maintainers' \texttt{0.30.3}. An age
cutoff instead exempts the old popular backports, and a strict one-day
cooldown alone keeps out \DynIgnoreAfterCooldownKeptOutOneDay{} of the
\DynIgnoreAfterPosReleases{} reconstructed malicious releases. A declared
policy makes this trade-off an explicit, public choice for each line:
declaring lines as published by hand admits \DynDeclReplayEnforcingIncidentPerLineAdmitWithEarlierEvidence{}
or~\DynDeclReplayAsReplayedIncidentPerLineAdmitWithEarlierEvidence{} of
the \DynDeclReplayIncidentWithEarlierEvidence{} malicious releases that
pnpm flags (\S\ref{sec:discussion}). Pins on the full signing identity
alert on \DynSignerPooledAdjacentFullSanShare{} of packages with at least
two keyless-signed releases, largely because certificates name the Git
branch or tag (\S\ref{sec:rq3}).

This paper makes three contributions:
\begin{itemize}
  \item We measure pnpm's rule in three pnpm releases on two npm
  populations and against documented attacks, and trace most popular
  failures to one security backport (\S\ref{sec:rq1},
  \S\ref{sec:case-studies}).
  \item We show how history scope, release age, and evidence ranking
  control the rule, and replay two declared policies on the
  same releases (\S\ref{sec:rq2}).
  \item We compare the evidence that these five registries expose and the
  identity pins it supports (\S\ref{sec:rq3}), and derive recommendations
  for consumers, maintainers, and registries (\S\ref{sec:discussion}).
\end{itemize}

%% file: sections/background.tex
\section{Background and Threat Model}\label{sec:model}

\subsection{Publication Evidence}\label{sec:evidence}
A release reaches its registry through a \emph{release pipeline}: a
maintainer publishing by hand, or a CI workflow publishing for the
project. Consumer tools read the release's publication evidence from the
registry's public endpoints.
Table~\ref{tab:dynamics-evidence-registry} lists the endpoints we query,
their fields, and the share of sampled packages with each field on any
release. It counts packages because three packages hold
\DynEvidNpmTopThreeShareMin{} to \DynEvidNpmTopThreeShareMax{} of the npm
releases in each of samples~1--3.

npm and PyPI expose the most evidence. Both support \emph{trusted
publishing}~\cite{npm_trusted_publishers,pypi_trusted_publishers}: a CI
workflow registered as the package's \emph{trusted publisher} publishes
with its OpenID Connect (OIDC) identity instead of a maintainer's
credentials. A release can also carry an \emph{attestation}, a signed
record of the CI workflow that produced it: \emph{provenance} on
npm~\cite{npm_provenance} and a publish attestation on
PyPI~\cite{pep740}, which Table~\ref{tab:dynamics-evidence-registry}
lists under Provenance. The attestation's signature is \emph{keyless}: a
short-lived Sigstore certificate bound to the workflow's OIDC identity
replaces a long-lived key~\cite{newman2022sigstore}. The
certificate's Subject Alternative Name (SAN) names the signer: the
repository, workflow file, and Git ref (branch or tag), as in
{\urlstyle{tt}\nolinkurl{https://github.com/nodejs/undici/.github/workflows/release.yml@refs/heads/main}}.
When a project calls a reusable workflow from another repository, the
SAN names that repository. A separate field records the source
repository~\cite{fulcio_oid}; the two fields differ for
\DynSignerFOneSourceDiffReleases{} releases of
\DynSignerFOneSourceDiffPackages{} package in sample~1. npm also names
each release's publishing account and signs it with a registry
key~\cite{npm_registry_signatures}; PyPI names a publisher only
for uploads that carry an attestation.

\input{generated/tab_evidence_registry}

Maven Central requires a PGP signature on every
file~\cite{sonatype2026requirements}. The endpoints we query name the
project's registered namespace but no
uploader~\cite{sonatype2026namespace}. crates.io names the publishing
account and supports trusted publishing~\cite{crates2026trusted}, which
records the repository, run, and commit of an upload. The RubyGems
versions API returns no field that the rules of \S\ref{sec:rules}
compare. Its attestations API~\cite{rubygems2026api}
and Maven Central's \texttt{.sigstore.json}
files~\cite{sonatype2025sigstoreportal} serve keyless signatures
(\S\ref{sec:collection}).

\subsection{Consumer Rules}\label{sec:rules}
We study pnpm's rule (R1) and three identity pins (R2--R4). All four
compare a release's evidence with that of earlier releases, so earlier
releases set what later ones must show; SLSA calls this trust on first
use~\cite{slsa2025v12}. The pins compare the name
of a signer, key, or publisher. An \emph{alert}
is a release that a replayed rule reports. A \emph{block} is pnpm's
refusal to install a release; a \emph{flag} is either.

\paragraph*{\hspace{-\parindent}\bf R1: pnpm's no-downgrade policy.} In
pnpm~10, the policy is off by default~\cite{pnpm_trust_policy}. It ranks
a trusted publisher above provenance and provenance above no evidence.
We call these inputs \emph{trust evidence}; the term says nothing about
whether a release is safe. A version's \emph{history} holds every version
of the package published before it, whatever its version number; since
10.24.0, the history of a stable version leaves out
prereleases~\cite{pnpm_prerelease_fix}. pnpm blocks a version whose
evidence ranks below the strongest evidence in its history. In
pnpm~\DynPnpmVersion{}, a block stops the entire install with an error;
the check is unchanged through 10.34.6. pnpm~10 checks a version only
when it resolves a dependency to it, so a full install and a
frozen-lockfile install keep a version that the lockfile records, even
one that the policy would block. A lockfile-only run on an empty store
checks that version again when the manifest has changed. Three settings
limit the policy; pnpm~10 sets none by default. An
\emph{exemption} (\texttt{trustPolicyExclude}) skips a package or exact
versions. An \emph{age cutoff} (\texttt{trustPolicyIgnoreAfter}, since
\DynPnpmIgnoreAfterSince{}) skips releases older than the cutoff. A
\emph{cooldown} (\texttt{minimumReleaseAge}) holds back releases younger
than a set age, whatever their evidence, as Dependabot's
does~\cite{tanaka2026cooldown}. On sampled npm, PyPI, and crates.io
histories, R1 uses a proxy rank built from keyless signature, workflow
identity, and attestation; for npm, we also emulate pnpm's rule
(\S\ref{sec:pnpm-rule}).

\paragraph*{\hspace{-\parindent}\bf pnpm 11 and 12.} Later majors keep the
policy opt-in and still order history by publish
time~\cite{pnpm_trust_policy,pnpm_releases}. Since
\DynPnpmCooldownDefaultSince{}, a one-day cooldown is the default. This
default is loose: it allows a younger release when no match is old
enough. Since \DynPnpmLockfileVerifySince{}, an explicitly set cooldown
is \emph{strict} and allows no younger release, and installs
re-check the lockfile and abort on a failure unless
\texttt{trustLockfile} (since \DynPnpmTrustLockfileSince{}) is set. Since \DynPnpmTrustedPublisherNeedsProvenanceSince{}, a
trusted publisher counts only alongside provenance. Since
\DynStagedRankSince{}, a \emph{staged publish}, approved with a second
factor, ranks above it; pnpm added this rank after a report of blocked
staged releases~\cite{pnpm_issue_11887}. Since
\DynPnpmExcludePruneSince{}, an opt-in setting removes unused
exemptions. Since 11.28 and 12.7, a newly resolved range selects the
newest match that passes the check. Blocked exact versions and ranges
with no passing match still fail.

\paragraph*{\hspace{-\parindent}\bf R2--R4: identity pins.} R2, the
keyless signer pin, compares certificate SANs. SLSA
describes such comparisons of provenance across versions and allows safe
branches or tags to be treated as benign~\cite{slsa2025v12}. R3, the
key-id pin, compares the identifier of the signing key, such as npm's
registry key or a Maven PGP key or subkey. R4, the publisher pin,
compares the publisher that the registry names, an account or a trusted
publisher. On npm, it ignores the workflow identity in provenance;
crates.io names no account for a trusted-publisher upload, so R4 skips
those releases. Commercial scanners raise similar alerts for a new
publisher~\cite{socket_alerts}. A pin alerts on a change only when both
releases carry a value. Gradle, cosign, and \texttt{gh} instead check
keys, checksums, or identities that the consumer
lists~\cite{gradle_verification,cosign_verify,gh_attestation_verify},
and PEP~751 lock files can record a publisher~\cite{pep751}.

\begin{figure}[t]
\centering
\input{figures/fig_running_example}
\caption{Running example: \texttt{undici-types}. After one release
drops provenance, pnpm~\DynPnpmVersion{} blocks every later release
without it, including \texttt{7.0.0} on a new line. A per-line history
passes \texttt{7.0.0}, and adjacent memory flags only the two releases
that drop provenance. Triangles mark the versions to which
\texttt{@types/node} releases resolve.}
\Description{A table with one row per release of the package, from 6.18.2 to 7.2.1. Columns show whether the release has provenance, whether it is a version that @types/node resolves to, and whether the adjacent-memory rule, the pnpm rule and the per-line rule pass or block it. The row for 7.0.0 is set in bold.}
\label{fig:running-example}
\end{figure}

\paragraph*{\hspace{-\parindent}\bf Design choices.} The \emph{history
scope} sets which earlier releases a rule compares with: all of them,
those in the same \emph{channel} (stable or prerelease, as marked by
version syntax), or those on the same \emph{maintenance line} (major
version, or major and minor for 0.x versions), a \emph{per-line}
history. The most recently published release in that history is the
\emph{predecessor}; with no restriction, it is the \emph{timestamp
predecessor}. A \emph{backport}, a release on an older line published
after releases of a newer line, can thus have a timestamp predecessor on
a newer line. \emph{Memory} sets how much of the history a rule uses.
\emph{Adjacent} memory compares a release only with its predecessor.
Under \emph{history} memory, a pin alerts only on values absent from the
history, and R1 compares with the strongest evidence in it, as pnpm does.
R2 compares the full SAN, accepts any tag ref, removes refs, or compares
only the signing workflow's repository.

\subsection{Threat Model}\label{sec:threat-model}

\paragraph*{\hspace{-\parindent}\bf Attacker.} We consider an attacker who
publishes a release with a maintainer's stolen publishing token or
registry account, as in the incidents of \S\ref{sec:introduction}. The
attacker bypasses the project's CI workflow, so the release can lack
evidence that earlier CI releases carried, such as a trusted publisher or
provenance. The rules can detect such a difference only when comparable
evidence exists in the history.

\paragraph*{\hspace{-\parindent}\bf Benign differences.} The same
comparisons can flag ordinary releases: a certificate identity can name
the release's own tag, and a package can publish through several
pipelines at once. An alert therefore marks an evidence change for a
person to review. It does not prove compromise or authorization.

\paragraph*{\hspace{-\parindent}\bf Out of scope.} An attacker who
controls the project's CI workflow can publish with the usual evidence
and is outside this threat model.

\subsection{Running Example}\label{sec:running-example}
Figure~\ref{fig:running-example} follows \texttt{undici-types}, on which
\texttt{@types/node} depends. The releases shown before \texttt{6.19.3}
carry provenance. On July~22, 2024, \texttt{6.19.3} appeared without it.
npm names the same publishing account before and after; only the
provenance is missing. Releases through \texttt{7.0.0}, and later
\texttt{7.2.0}, also lack provenance. All rank below the provenance of
earlier releases, so pnpm blocks them, including \texttt{7.0.0} on a new
line (Finding~3). In our runs, the 20.x and 22.x releases of
\texttt{@types/node} require \texttt{undici-types}
\texttt{\textasciitilde{}6.21.0}, which only the blocked \texttt{6.21.0}
satisfies.

%% file: generated/tab_evidence_registry.tex
\begin{table}[t]\centering\footnotesize
\caption{Evidence per registry in the endpoints we query. npm and PyPI return workflows, provenance, and keyless signatures; Maven PGP keys; crates.io trusted publishing for a minority of packages; RubyGems no publisher.}\label{tab:dynamics-evidence-registry}
\setlength{\tabcolsep}{3.5pt}
\begin{tabular}{@{}lcccccc@{}}\toprule
 & \multicolumn{2}{c}{Identity} & \multicolumn{2}{c}{CI build} & \multicolumn{2}{c}{Signature} \\
\cmidrule(lr){2-3}\cmidrule(lr){4-5}\cmidrule(l){6-7}
Registry & Publisher & Repo & Workflow & Provenance & Key & Keyless \\\midrule
npm & 100 & 86 & 42 & 43 & 100 & 43 \\
\multicolumn{5}{@{\hspace{0.8em}}l}{\textit{packument, attestations API, deps.dev}} & \multicolumn{2}{r@{}}{\llap{rules: R1--R4}} \\[2pt]
PyPI & 40\rlap{\textsuperscript{b}} & 87 & 37 & 40 & -- & 40 \\
\multicolumn{5}{@{\hspace{0.8em}}l}{\textit{JSON API, Simple index, Integrity API, deps.dev}} & \multicolumn{2}{r@{}}{\llap{rules: R1, R2, R4}} \\[2pt]
Maven & 100\rlap{\textsuperscript{a}} & 91 & -- & -- & 99 & -- \\
\multicolumn{5}{@{\hspace{0.8em}}l}{\textit{directory listing, POM, POM signature}} & \multicolumn{2}{r@{}}{\llap{rules: R3\textsuperscript{d}}} \\[2pt]
crates.io & 100 & 94 & -- & 13\rlap{\textsuperscript{c}} & \multicolumn{2}{c}{checksum only} \\
\multicolumn{5}{@{\hspace{0.8em}}l}{\textit{versions API}} & \multicolumn{2}{r@{}}{\llap{rules: R1, R4}} \\[2pt]
RubyGems & -- & 63 & -- & -- & \multicolumn{2}{c}{checksum only} \\
\multicolumn{5}{@{\hspace{0.8em}}l}{\textit{versions API}} & \multicolumn{2}{r@{}}{\llap{rules: none\textsuperscript{d}}} \\
\bottomrule\end{tabular}
\par\smallskip\parbox{\columnwidth}{\scriptsize\raggedright Percent of packages in samples 1--3 with the field on any release, ``--'': never populated. Italic: Web endpoints queried; rules: those whose input varies within a package. \textsuperscript{a}registered namespace (groupId); \textsuperscript{b}trusted publisher from the attestation; \textsuperscript{c}trusted-publishing record (repository, run, commit), no attestation; \textsuperscript{d}not in these records: the RubyGems attestations API and Maven Central \texttt{.sigstore.json} files, queried separately in October 2026, return keyless signatures for 5.9\% and 0.1\% of the releases (6.1\% and ${<}0.1$\% over samples 1--4).}
\end{table}

%% file: figures/fig_running_example.tex
% Native-size TikZ diagram generated from the canonical running-example CSV.
\captionsetup{skip=4pt}
\input{figures/publication/running_example_tikz}%

%% file: sections/study_design.tex
\section{Methods}\label{sec:design}

\subsection{pnpm on Two npm Populations}\label{sec:ecosystem}

\paragraph*{\hspace{-\parindent}\bf Populations.} Both populations come from the Ecosyste.ms
package index. The \emph{popular} population holds all
\DynEcoTopTargets{} most-downloaded npm packages. The \emph{recently
updated} one holds \DynEcoRandomTargets{} packages drawn at random,
excluding the popular ones, from a snapshot of the
\DynEcoRandomPoolListed{} most recently updated packages, whose latest
releases span \DynEcoRandomPoolWindowHours{} hours. Of its
packages, \DynEcoRandomScopedShare{} carry one of \DynEcoRandomScopes{} npm
scopes, and \DynEcoRandomZeroDownloadShare{} had no downloads in the past
month. A second draw of \DynEcoDayTwoTargets{} recently updated packages
on \DynEcoDayTwoDateProse{}, by the same procedure, tests whether its
failure rate depends on the day.

\paragraph*{\hspace{-\parindent}\bf Live runs.}\label{sec:pnpm-rule} On September~27, 2026, we
resolved every package's latest version against the live npm registry
with pnpm~\DynEcoPnpmVersion{} and \texttt{trustPolicy} set to
\texttt{no-downgrade}, leaving its other settings at their defaults: no
exemption, cooldown, or age cutoff. Each run starts without a lockfile and only
writes one, resolving every version with install scripts disabled. We report
pnpm~10 first because each of its blocks fails the install. Three further
sets of runs vary one factor each.
On \DynPnpmCurrentObservedDateProse{}, we ran
pnpm~\DynPnpmMidVersion{} and pnpm~\DynPnpmCurrentVersion{}, the latest
on npm that day, on the same package versions; we ran the latter with
and without its default cooldown, and pnpm~\DynPnpmCurrentLegacyVersion{}
as a same-day control. Comparing the lockfiles that
pnpm~\DynPnpmCurrentVersion{} writes with and without the policy shows
which resolved versions the policy changes.
\label{sec:ignore-after} On September~29, we ran both populations
with pnpm~\DynIgnoreAfterPnpmVersion{} and age cutoffs of 1, 7, 30, and
365 days. On
\DynEcoMostDownloadedObservedDateProse{}, we ran
pnpm~\DynEcoMostDownloadedVersion{} on the version of each popular
package with the most downloads in the preceding week, as reported by
npm's downloads API, and again on its latest version.

\paragraph*{\hspace{-\parindent}\bf Blocking releases.} pnpm's error names the release that it
blocks, the \emph{blocking release}. A control run without the policy
succeeds for \DynEcoTopControlConfirms{} of the \DynEcoTopFailing{}
failing popular packages, \DynEcoRandomControlConfirmed{} of the
\DynEcoRandomFailing{} recently updated ones, and all
\DynEcoMostDownloadedControlConfirms{} that fail at the most-downloaded
version; \DynEcoRandomUnrelatedFailures{} recently updated installs that
failed for other reasons count only in the denominator. One blocking
release can break many packages, so we compute the interval of the
recently updated failure rate by resampling blocking releases and also
count each npm scope once. A blocking release is
a \emph{backport} if its package had already published a stable release
on a higher major line. It is a \emph{security backport} if it is also
the first patched version of a GitHub advisory.

\paragraph*{\hspace{-\parindent}\bf Emulation.} Removed releases and sampled histories require
offline replay, so we emulate pnpm's rule on their archived evidence. The
emulation matches pnpm's compiled check on all
\DynPnpmEmulationComparedVersions{} npm versions captured in
samples~1--3 (\S\ref{sec:samples}), and live pnpm~\DynPnpmVersion{} agrees with it on
\DynPnpmConcordanceAgree{} of \DynPnpmConcordanceCases{} cases: every
release there that it blocks and \DynPnpmSampleSize{} randomly drawn stable
releases that it passes.

\paragraph*{\hspace{-\parindent}\bf Configurations.} We searched GitHub on September~29 for public
pnpm configurations that enable no-downgrade and recorded their
exemptions, age cutoffs, and cooldowns. On October~4, we collected their
root \texttt{package.json} files, which name the pnpm version that the
project uses and may declare one of \DynAdoptionWebFrameworkFrameworks{}
Web UI frameworks, such as React, Vue, or Next. These versions motivate
the comparison across pnpm releases: of \DynPnpmMidPinsExact{} files that
name an exact version, \DynPnpmMidPinsOriginalShare{} name pnpm~10 or
earlier, \DynPnpmMidPinsLaterScaleShare{} a release with the later
ranking that still fails on every block, and
\DynPnpmMidPinsOlderVersionShare{} a release that can resolve an older
version; \DynPnpmMidPinsEarlyElevenShare{} name pnpm~11.0--11.4, which we
did not run.

\subsection{Incident Replay and Advisory Census}\label{sec:positive-controls}

\paragraph*{\hspace{-\parindent}\bf Incidents.} An \emph{incident} is one documented attack in
which an attacker published malicious versions of existing packages with a
maintainer's stolen credentials or publishing token. We reconstruct
\DynPosIncidents{} incidents, \DynPosIncidentsNpm{} on npm and
\DynPosIncidentsPypi{} on PyPI, with \DynPosMalReleases{} malicious
releases of \DynPosPackages{} packages, from \texttt{eslint-scope} in 2018
to \texttt{axios} in
2026~\cite{axios2026postmortem,posthog2025postmortem,pypi2026litellm_telnyx}.
An incident qualifies when an advisory or postmortem names the malicious
versions and attributes them to stolen credentials or tokens. We collected
these incidents from public write-ups and took one or two packages from
campaigns that hit many, such as the Shai-Hulud worms. In
\DynPosAccessTokenIncidents{} incidents the attacker held only a publishing
token, and in \DynPosAccessAccountIncidents{} the maintainer's registry
account. Registries no longer serve these releases, so we reconstruct
their evidence, publisher, and removal time from advisories and
postmortems and mark inferred values (Table~\ref{tab:positive-controls}).
We insert these records into the captured histories to replay pnpm's rule
and R1--R4. Benign releases within \DynPosWindowDays{} days before or
after each incident measure false alarms.

\paragraph*{\hspace{-\parindent}\bf Census.} An advisory census extends coverage to every npm
malware advisory in the GitHub Advisory Database from April 2023 to
September~27, 2026. It counts a package as \emph{attacked} if at least
\DynCensusMinPrior{} of its versions that no advisory lists appeared at
least \DynCensusPriorGapDays{} days before the first malicious one, and
if npm took down at least one malicious version. A \emph{wave} groups at
least \DynCensusWaveMinPackages{} packages whose first malicious versions
follow one another within \DynCensusWaveGapHours{} hours, as in a worm. A
census incident is a wave or a package outside any wave. Because one wave
can hold many packages of one project, we report shares of both packages
and incidents.

\subsection{Release Histories from Five Registries}\label{sec:samples}

\paragraph*{\hspace{-\parindent}\bf Samples.} Our four samples cover the five package registries
without sharing packages (Table~\ref{tab:dynamics-samples},
Appendix~\ref{app:samples}). Sample~1 draws packages from each registry's
public activity feed, stratified by publishing activity, so it
over-represents frequent, often automated publishers. Samples~2--4 draw
from the Ecosyste.ms package index and target 30, 40, and 50 packages per
stratum. Their \emph{popular} stratum comes from each registry's
\DynFramesFTwoNpmPopularPool{} most-downloaded packages
(most-depended-upon for Maven Central). Their \emph{random} stratum draws
from \DynFramesFTwoNpmRandomPool{} candidates on random pages of the
index's default listing. Its order is undocumented, so the draw may not
be uniform. Each sample excludes the packages of earlier samples and
those used to develop the analysis, including many popular ones such as
\texttt{semver}: in sample~2, \DynFramesPopularExcludedMinFTwo{} to
\DynFramesPopularExcludedMaxFTwo{} of a registry's popular candidates.
Sample~4 excludes code repositories from samples~1--3.

\paragraph*{\hspace{-\parindent}\bf Eligibility.} A package needs at least four releases
between April~6, 2024 and September~2, 2026, whatever their publication
evidence. In sample~2, \DynFramesFTwoPopularPassRate{} of the checked
popular candidates and \DynFramesFTwoRandomPassRate{} of the random ones
pass, so the samples describe packages that release regularly.

\paragraph*{\hspace{-\parindent}\bf Collection.}\label{sec:collection} We captured every release
in each sample from April~6, 2024 through its collection date. The
collector queried the endpoints of Table~\ref{tab:dynamics-evidence-registry} on
September~6, 2026 for sample~1, September~26 for samples~2 and~3, and
\DynFFourCollectionDateProse{} for sample~4. Samples~1--4 hold
\DynFOneCoordinates{}, \DynFTwoCoordinates{}, \DynFThreeCoordinates{}, and
\DynFFourPipePackages{} packages. On PyPI, the collector reads the
attestations of at most \DynPypiFilesFileCap{} files per release;
\DynPypiFilesMixedReleases{} of the \DynPypiFilesAttestedReleases{}
attested releases also holds a file without one. On
\DynSupplCollectedOnProse{}, we also fetched the RubyGems attestations
API and Maven Central's \texttt{.sigstore.json} files for every sampled
release of these registries, the KEYS files in which Apache projects list
their signing keys~\cite{apache_release_distribution}, and the crates.io
setting that refuses token uploads~\cite{crates2026trustpubonly}.

\paragraph*{\hspace{-\parindent}\bf Rule replay.}\label{sec:design-rules}
We replay the rules of \S\ref{sec:rules} in every sample under each
history scope and memory, and R2 under each identity it can compare
(Table~\ref{tab:dynamics-replication}). For signer pins, we count the
packages that alert among those with at least two keyless-signed
releases. A variant of R4 also compares the publishing account's e-mail
address. Intervals are 95\%: Wilson for shares of packages and
Newcombe for differences between strata. Packages from one code repository
share pipelines, so we resample repositories for signer-pin shares.

\paragraph*{\hspace{-\parindent}\bf Alert decomposition.}\label{sec:attribution} From each
rule's \emph{naive variant} (timestamp predecessor, adjacent memory, full
identity), five mechanisms relax the comparison: they ignore
registry-wide events, stay within a channel or line, ignore recurring
changes, and drop the certificate's ref
(Appendix~\ref{app:waterfall-orders}). Alerts that no mechanism removes
are \emph{residual}.

\paragraph*{\hspace{-\parindent}\bf Declared policies.}\label{sec:declared} We replay two
policies that a package could declare at its registry, inferring each
declaration from its earlier releases. The \emph{registered-pipeline} policy accepts a release
only if it carries evidence from a repository that the package's earlier
evidence names; it compares no ranks. The \emph{per-line declaration}
marks a maintenance line \emph{manual} when none of its earlier releases
carries evidence; a new line inherits the package's history. It accepts
every release on a manual line, and other lines require evidence from the
line's repositories. Two conventions bound the result. \emph{As
replayed}, a release's history keeps the releases that the policy would
have refused. \emph{Enforced}, as if the registry had applied the
policy, a refused release never enters a later history, and the
registered pipeline applies from the package's first release with
evidence. We apply both
policies to the blocking releases of \S\ref{sec:ecosystem} and the
malicious releases of \S\ref{sec:positive-controls}, and report both
conventions where they differ.

\subsection{Validation and Replication}\label{sec:validity}
On sample~1, an independently written parser re-reads the responses of
\DynValRefReleases{} random releases and agrees with our records on every
field it can read (\DynValRefChecks{} field checks). OpenSSL
re-parses \DynValCertChecked{} signing certificates and recovers the
same issuers and identities. Some repositories,
such as typescript-eslint's monorepo, appear in several samples, so
agreement between samples could reflect shared projects. Sample~4
therefore shares no repository with the others and keeps the same record
fields, rules, and eligibility (Table~\ref{tab:dynamics-replication}).
After collecting the samples, we defined the package shares and their
two-release denominator, the variant that accepts any tag, the pooling of
samples~2--4, the classification of failing packages by role, and the
declared policies with their conventions.
We will release the stored responses, the pnpm run outputs, and the
analysis code with the final version of the paper.

%% file: sections/results.tex
\section{Results}\label{sec:results}

We answer the three research questions in order:
\begin{enumerate*}[label=\textbf{RQ\arabic*:}]
  \item What does pnpm's policy cost, and which attacks does it flag?
  \item Which design choices control these outcomes, and what would a
  declared policy change?
  \item What evidence do package registries expose, and how do identity pins
  behave?
\end{enumerate*}

\subsection{RQ1: What Does pnpm's Policy Cost, and Which Attacks Does It
Flag?}\label{sec:rq1}

The live runs (\S\ref{sec:ecosystem}) measure what the policy costs; the
census and the reconstructed incidents (\S\ref{sec:positive-controls})
show which attacks it flags.

\input{generated/tab_positive_controls}

\paragraph*{\hspace{-\parindent}\bf Finding 1: a few backports cause most failed installs.} With \texttt{trustPolicy} set to \texttt{no-downgrade}
and other settings at their defaults, pnpm~\DynEcoPnpmVersion{} cannot
install the latest version of \DynEcoTopFailShare{} of the popular npm
packages (\DynEcoTopFailing{} of \DynEcoTopTargets{}) and of
\DynEcoRandomFailShare{} of the recently updated ones
(\DynEcoRandomFailing{} of \DynEcoRandomTargets{}).
Resampling blocking releases gives the two shares 95\% intervals of
\DynEcoTopFailShareLo{}--\DynEcoTopFailShareHi{} and
\DynEcoRandomFailShareLo{}--\DynEcoRandomFailShareHi{}, and counting each
npm scope once gives \DynEcoRandomScopeUnitsFailShare{}. In the second
draw, a week later, \DynEcoDayTwoFailShare{} of installs fail
(\DynEcoDayTwoFailShareLo{}--\DynEcoDayTwoFailShareHi{}), and
\DynEcoDayTwoFailingOnDayOneBlockersShare{} of these failures name a
blocking release of the first draw.

Dependencies spread these blocks to other packages. The blocked release
belongs to a dependency in \DynEcoTopTransitiveBlocks{} of the
\DynEcoTopFailing{} popular failures and \DynEcoRandomTransitiveBlocks{} of
the \DynEcoRandomFailing{} recently updated ones. All
\DynEcoTopFailingRoleTooling{} failing popular packages are JavaScript
build, lint, or test tools: \DynEcoTopFailingRoleBuild{} Babel plugins,
\DynEcoTopFailingRoleLint{} ESLint packages, and
\DynEcoTopFailingRoleTest{} test packages such as Jest's.

The \DynEcoTotalFailing{} failures of the two populations trace to
\DynEcoDistinctBlockers{} blocking releases. Of these,
\DynEcoBlockerNoEvidenceReleases{} carry neither a trusted publisher nor
provenance; the other \DynEcoBlockerProvenanceAfterTpReleases{} carry
provenance, which ranks below the trusted publisher of an earlier release.
The \DynEcoOlderLineBackportReleases{} blocking releases that are backports
to an older major line cause \DynEcoTopOlderLineBackportFailing{} of the
popular failures and \DynEcoRandomOlderLineBackportFailing{} of the
recently updated ones, and \DynEcoSecurityBackportReleases{} of these
backports fix published vulnerabilities. One of them,
\texttt{semver}~\texttt{6.3.1}, alone causes \DynEcoTopTopBlockerFailing{} of the
\DynEcoTopFailing{} popular failures (\S\ref{sec:case-studies}) and
\DynPnpmCurrentDefaultTopFailingOnTopBlocker{} of the
\DynPnpmCurrentDefaultTopFailing{} under pnpm~\DynPnpmCurrentVersion{}.

These installs request the latest version, which is also the
most-downloaded one for \DynEcoMostDownloadedIsLatest{} of the
\DynEcoMostDownloadedCovered{} popular packages. Requests for the
most-downloaded version fail more often: \DynEcoMostDownloadedFailing{}
installs, against \DynEcoMostDownloadedFailingLatest{} for the latest
version on the same day. \texttt{semver}~\texttt{6.3.1} is its package's
most-downloaded version and fails when requested directly.

Of the \DynAdoptionHeadlineRepos{} public GitHub repositories that we
found enabling the policy, \DynAdoptionHeadlineWithExcludeShare{} exempt
releases from it, mostly by exact version. The most exempted package is
\texttt{semver}, in
\DynAdoptionTopOnePackages{} repositories, \DynAdoptionSemverSixThreeOne{}
of which name \texttt{6.3.1}, the backport that causes most failed popular
installs (\S\ref{sec:case-studies}).

\paragraph*{\hspace{-\parindent}\bf Finding 2:
\DynCensusAllPriorEvidenceShare{} of attacked packages had earlier
evidence, with a 95\% interval of
\DynCensusIncidentBootPriorEvidenceLo{}--\DynCensusIncidentBootPriorEvidenceHi{}
when incidents are resampled.} The policy can flag an attack only if a
package's earlier releases set a higher evidence rank. In the advisory census,
\DynCensusAllPriorEvidence{} of the \DynCensusAllPackages{} attacked
packages had provenance or a trusted publisher on a release before the
attack. The packages come from \DynCensusAllIncidents{} incidents,
\DynCensusAllIncidentsPriorEvidence{} of them with earlier evidence, and
the package share depends on which campaigns the census includes: across the
\DynCensusByWaveLargestCount{} largest of its \DynCensusByWaveWaves{}
waves, it ranges from \DynCensusByWaveLargestShareMin{} to
\DynCensusByWaveLargestShareMax{}, and two waves supply
\DynCensusTopTwoPriorEvidenceShare{} of the packages with earlier evidence
and \DynCensusTopTwoMalFlaggedShare{} of the flagged versions; without
them, the share is \DynCensusWithoutTopTwoPriorEvidenceShare{}. The
interval in the heading therefore resamples incidents; a wave window of
\DynCensusWindowDayHours{} hours or one week changes it to
\DynCensusWindowDayPriorEvidenceLo{}--\DynCensusWindowDayPriorEvidenceHi{}
or
\DynCensusWindowWeekPriorEvidenceLo{}--\DynCensusWindowWeekPriorEvidenceHi{}.
pnpm's rule flags \DynCensusAllMalFlagged{} of the
\DynCensusAllMalVersions{} malicious versions
(\DynCensusAllMalFlaggedShare{}), an upper bound because we assume that
the versions npm removed lacked evidence.
Attackers also published
\DynCensusMalEvidenceVersions{} malicious versions of
\DynCensusMalEvidencePackages{} packages through the project's own trusted
publisher. The rule flags none of them; they illustrate the
workflow compromise outside our threat model (\S\ref{sec:threat-model}).

The reconstructed incidents show the same dependence
(Table~\ref{tab:positive-controls}). The rule flags malicious releases in
\DynPosDeployedCaughtIncidents{} of the \DynPosIncidents{} incidents,
exactly those whose package had published provenance or attestations
before. With history memory,
as in pnpm, it flags \DynPosROneHistoryCaught{} of the
\DynPosMalReleases{} malicious releases; adjacent memory flags only
\DynPosROneAdjacentCaught{}. This gap follows from our
reconstruction, in which malicious releases carry no evidence; in
\DynPosDeployedCaughtInferredEvidence{} of the
\DynPosDeployedCaughtIncidents{} flagged incidents, that absence is
inferred ($\ddagger$). pnpm's ranks test only whether evidence is
present, so with the access each incident shows, the attackers could have
passed \DynPosBypassWithAccessReleases{} of the
\DynPosBypassFlaggedReleases{} flagged releases by adding evidence:
\DynPosBypassProvenanceReleases{} follow histories with provenance
alone, which provenance from a repository the attacker controls matches,
and \DynPosBypassTrustedPublisherReleases{} follow a trusted publisher
that the attacker, who held the account, could replace. Within \DynPosWindowDays{} days of the incidents,
the rule also flags \DynPosROneHistoryBenign{} benign releases,
\DynPosBenignFlagLitellmPypi{} of them \texttt{litellm} releases without
the attestations that the project adopted after its incident. The census
holds \DynPosNotReplayedHeadlineIncidents{} further incidents for which a
source states or implies stolen credentials, and the rule flags none of
their \DynPosNotReplayedHeadlineMalVersions{} malicious versions.

\paragraph*{\hspace{-\parindent}\bf Identity pins.} No
reconstructed malicious release is keyless-signed, and pins skip missing
evidence. The publisher pin with history memory passes all
\DynPosMalPublisherRecovered{} malicious npm releases whose publisher we
identified (\DynPosMalPublisherInferred{} inferred, $\dagger$), because
the attacker uses a past publisher's account.

\paragraph*{\hspace{-\parindent}\bf Cooldown.} The reconstructed
malicious releases stayed online for a median of
\DynIgnoreAfterPosMedianLiveHours{} hours and, where known, at most
\DynIgnoreAfterPosMaxLiveDays{} days. A strict one-day cooldown
(\S\ref{sec:rules}) therefore keeps out
\DynIgnoreAfterCooldownKeptOutOneDay{} of the
\DynIgnoreAfterPosReleases{} releases, whatever their evidence
(Table~\ref{tab:ignore-after}, Appendix~\ref{app:pnpm-blocks}). The other
\DynIgnoreAfterCooldownOutliveOneDay{} belong to
\texttt{eslint-config-prettier} and \texttt{@rspack/core};
\DynIgnoreAfterCooldownOutliveUnknownRemovalOneDay{} has no known removal
time. The policy flags all
\DynIgnoreAfterCooldownFlaggedOutliveOneDay{}, and the cooldown alone
keeps out \DynIgnoreAfterCooldownFlaggedKeptOutOneDay{} of the
\DynIgnoreAfterPosFlagged{} releases that the policy flags; a seven-day cooldown keeps out
\DynIgnoreAfterCooldownKeptOutSevenDays{} of all \DynIgnoreAfterPosReleases{}. For consumers, the policy adds to a strict cooldown only when a malicious
release outlives the cooldown and carries weaker evidence than an earlier
release.

\subsection{RQ2: Which Design Choices Control These Outcomes, and What
Would a Declared Policy Change?}\label{sec:rq2}

We vary three design choices of the rule and replay the two declared
policies (\S\ref{sec:declared}) on the releases of RQ1.

\input{generated/tab_declared_policies}

\paragraph*{\hspace{-\parindent}\bf Finding 3: per-line histories and
declarations pass backports and attacks on older lines.}
pnpm~\DynPnpmVersion{} keeps prereleases out of a stable release's history
but keeps one history across maintenance lines
(Table~\ref{tab:dynamics-history-scope}). In samples~\mbox{1--3}, it blocks
\DynPnpmInFrameStableBlocksTotal{} stable releases
(Table~\ref{tab:dynamics-pnpm-blocks}),
\DynPnpmInFrameStableBlocksTopPackage{} of them from one binary package,
\texttt{@landstrip/\allowbreak landstrip-darwin-arm64}; all
\DynPnpmBlockerNoEvidence{} carry neither a trusted publisher nor
provenance. One of them is the
security backport \texttt{uuid} version~\texttt{13.0.1}, which pnpm blocks because its
history spans lines (\S\ref{sec:case-studies}).

Table~\ref{tab:declared-policies} compares pnpm's rule with a per-line
history and the two declared policies on the releases of Findings~1
and~2. A per-line history passes \texttt{uuid}~\texttt{13.0.1} and the
\DynLineBackportsTotal{} security backports of Finding~1,
\DynLineBackportsPassPerLineVacuous{} of them only because no earlier
release on their line carries evidence. It removes at most
\DynDeclReplayAsReplayedEcoLineHistoryFailuresRemovedBoth{} of the
\DynDeclReplayEcoFailuresBoth{} failed installs: pnpm names one blocking
release per failed install, and admitting it may expose another blocked
dependency. It also passes the
attacker's \texttt{axios}~\texttt{0.30.4} along with the maintainers'
\texttt{0.30.3}, and the attackers'
\texttt{eslint-config-\allowbreak prettier} \texttt{8.10.1} and
\texttt{9.1.1}, so it flags \DynPosROneHistoryLineCaught{} of the
\DynPosROneHistoryCaught{} malicious releases that pnpm flags. The two
\texttt{axios} releases name the same account and carry no provenance, so
pnpm's rule and the name-based pins see the same inputs. They differ in
the account's e-mail address, which the attacker had changed. A publisher
pin that also compares this address flags both malicious \texttt{axios}
releases and raises \DynPubEmailHistoryAdded{} more alerts in
\DynPubEmailHistoryAddedPackages{} packages across the four samples, but
an attacker who controls the account can keep or restore the address.

\paragraph*{\hspace{-\parindent}\bf Registered pipeline.} No reconstructed
malicious release carries evidence, so a declared policy separates
malicious from benign releases only in packages that published evidence
before. There, the registered-pipeline policy refuses the same
\DynDeclReplayIncidentWithEarlierEvidence{} malicious releases as pnpm.
It also refuses \DynDeclReplayAsReplayedEcoRegisteredRefuse{} of the
\DynDeclReplayEcoReleases{} blocking releases
(\DynDeclReplayEnforcingEcoRegisteredRefuse{} when enforced), so a project
must publish its backports through the pipeline. Each of these refusals
comes from missing evidence, except
\DynDeclReplayEnforcingRepositoryDecided{} that the repository condition
decides when enforced. The other
\DynDeclReplayIncidentWithoutEarlierEvidence{} malicious releases belong
to \DynDeclReplayIncidentWithoutEarlierEvidencePackages{} packages with no
pipeline to declare;
\DynDeclReplayIncidentWithoutEarlierEvidenceBeforeNpmProvenance{} predate
npm provenance. As replayed, the policy refuses them because it
refuses every release without evidence, including
\DynDeclReplayWindowBenignNoEvidence{} of the \DynDeclReplayWindowBenign{}
benign releases in the incident windows
(Appendix~\ref{app:declared-samples}); enforced, it refuses none of the
\DynDeclReplayIncidentWithoutEarlierEvidence{}.

\paragraph*{\hspace{-\parindent}\bf Per-line declaration.} This policy
reproduces the trade-off of the per-line history. It admits
\DynDeclReplayAsReplayedIncidentPerLineAdmitWithEarlierEvidence{} of the
\DynDeclReplayIncidentWithEarlierEvidence{} malicious releases
(\DynDeclReplayEnforcingIncidentPerLineAdmitWithEarlierEvidence{} when
enforced) and removes at most
\DynDeclReplayAsReplayedEcoPerLineFailuresRemovedBoth{} of the
\DynDeclReplayEcoFailuresBoth{} failed installs
(\DynDeclReplayEnforcingEcoPerLineFailuresRemovedBoth{} when enforced).
For registries,
a declaration turns this exposure into an explicit choice that a project
can close line by line (\S\ref{sec:discussion}).

\paragraph*{\hspace{-\parindent}\bf Finding 4: an age cutoff exempts the popular blocking
releases.} With pnpm~\DynIgnoreAfterPnpmVersion{}, no popular
install fails at the cutoffs we ran, up to one year
(Table~\ref{tab:ignore-after}): every popular blocking release was at least
\DynIgnoreAfterTopMinBlockerAgeDays{} days old on September~29, including
\texttt{semver}~\texttt{6.3.1} from July 2023.
The blocking releases of recently updated packages are younger, so
\DynIgnoreAfterRandomFailingThirtyDays{} of their installs fail at 30 days,
against \DynIgnoreAfterRandomFailingNone{} with no cutoff on the same day.
The cutoff separates releases by age: it exempts old backports and attacks
that outlive it, and it still checks new backports and new attacks. In \DynAdoptionIgnoreAfterLeCooldown{} of the
\DynAdoptionCooldownWithIgnoreAfter{} enabling repositories that set both
a cutoff and a cooldown, the cutoff is no longer than the
cooldown, so the policy checks only packages that the cooldown exempts.
For tool authors, an age cutoff sets how long a release stays checked. It
does not show that an older release is safe.

\paragraph*{\hspace{-\parindent}\bf Finding 5: a finer ranking removes blocks and can add
others.} pnpm~11 changed the ranking twice (\S\ref{sec:rules}).
pnpm~\DynPnpmMidVersion{} has both changes and still fails on every block:
\DynPnpmMidTopFailing{} popular and \DynPnpmMidRandomFailing{}
recently updated installs, against \DynPnpmMidLegacyTopFailing{} and
\DynPnpmMidLegacyRandomFailing{} for pnpm~10 on the same day. Among the
recently updated installs that pnpm~10 fails,
\DynPnpmMidLegacyFailingRandomNowPass{} now complete and
\DynPnpmMidLegacyFailingRandomNowOther{} fails for another reason;
\DynPnpmMidLegacyPassingRandomNowFailing{} that pnpm~10 completes now
fail. No popular install changes. pnpm~\DynPnpmCurrentVersion{} shows
the same counts of installs that now pass and now fail
(Table~\ref{tab:pnpm-transition}, Appendix~\ref{app:pnpm-blocks}).

The installs that now pass had failed after an earlier release with a
trusted publisher but no provenance, which no longer counts as
evidence. Of the
\DynPnpmCurrentLegacyRandomPassingNowFailing{} new failures,
\DynPnpmCurrentNowFailingCauseStaged{} follow a staged publish that a
later release did not repeat. One staged publish can therefore raise the
rank required of later releases from the project's own CI. This result
rests on few installs. With its defaults, pnpm~\DynPnpmCurrentVersion{}
fails \DynPnpmCurrentDefaultTopFailing{} popular and
\DynPnpmCurrentDefaultRandomFailing{} recently updated installs; a further
\DynPnpmCurrentDefaultTopFallback{} and
\DynPnpmCurrentDefaultRandomFallback{} complete with an older version of
a dependency. For tool authors, each added rank is another level
below which ordinary releases can fall.

\subsection{RQ3: What Evidence Do Package Registries Expose, and How Do
Identity Pins Behave?}\label{sec:rq3}

We use the samples of \S\ref{sec:samples}. Unless noted, shares pool
samples~2--4, and Table~\ref{tab:dynamics-replication} reports each
sample.

\paragraph*{\hspace{-\parindent}\bf Finding 6: only npm exposes inputs for all four rules.}
Table~\ref{tab:dynamics-evidence-registry} maps each registry's fields to
the rules they support. Provenance concentrates in
popular packages:
\DynFramesPooledNpmPopularKeyless{} of popular and
\DynFramesPooledNpmRandomKeyless{} of random npm packages carry
keyless-signed provenance (difference \DynFramesPooledNpmKeylessDiff{}
percentage points, 95\% CI
\DynFramesPooledNpmKeylessDiffLow{}--\DynFramesPooledNpmKeylessDiffHigh{}).
Many packages publish on several channels or lines:
\DynPipeFTwoInterleavedPackages{} of sample~2's
\DynPipeFTwoPackages{} packages publish both prereleases and stable
releases, and \DynPipeFTwoStableBackportPackages{} publish stable
backports. npm's signing key, however, is the registry's own, so R3 on
npm alerts only when the registry rotates it.

Dropping evidence is rarer. Of the \DynProvLossPooledEver{}
packages in samples~2--4 whose stable releases ever carried provenance or
an attestation, \DynProvLossPooledLost{} later published a stable release
without it (\DynProvLossPooledShare{}, 95\% CI
\DynProvLossPooledShareLow{}--\DynProvLossPooledShareHigh{}). For tool
authors, even a single drop matters: pnpm blocks each later release
without evidence until it returns (\S\ref{sec:running-example}).

\paragraph*{\hspace{-\parindent}\bf Finding 7: signer pins on the full SAN alert on
\DynSignerPooledAdjacentFullSanShare{} of packages with at least two
keyless-signed releases.} R2 applies
SLSA's trust-on-first-use comparison to the certificate's
SAN~\cite{slsa2025v12}; we measure the effect of its Git ref. Of the
\DynSignerPooledTwoKeylessPackages{} pooled packages with at least two
keyless-signed releases,
\DynSignerPooledAdjacentFullSanPackages{} alert under a pin on the full
SAN; resampling their \DynSignerRepoRepositories{} repositories gives the
share a 95\% interval of
\DynSignerRepoAllFullSanShareLow{}--\DynSignerRepoAllFullSanShareHigh{}.
Removing the ref lowers the share to
\DynSignerPooledAdjacentSanWithoutRefShare{}
(\DynSignerPooledAdjacentSanWithoutRefPackages{} packages;
Table~\ref{tab:signer-pins}, Appendix~\ref{app:replication}). The default check of
\texttt{gh attestation verify} and PyPI's publisher record omit the ref, so
the full-SAN row is an upper bound on what deployed verifiers would
raise.

A package signing from release tags shows a new SAN per release, as \texttt{aiohttp} and \texttt{electron-to-\allowbreak chromium}
do. In the pooled samples,
\DynSignerRegPypiPooledTagOnlyShare{} of PyPI packages sign only from
tags, against \DynSignerRegNpmPooledTagOnlyShare{} on npm. Accepting any
tag therefore reduces alerts more for PyPI than for npm:
\DynSignerRepoNpmTagWildcardShare{} of
npm's packages and \DynSignerRepoPypiTagWildcardShare{} of PyPI's still
alert (difference \DynSignerRepoDiffTagWildcard{} percentage points,
95\% CI from resampling code repositories
\DynSignerRepoDiffTagWildcardLow{} to
\DynSignerRepoDiffTagWildcardHigh{}), against
\DynSignerRepoNpmFullSanShare{} and \DynSignerRepoPypiFullSanShare{} under
the full SAN (\DynSignerRepoDiffFullSan{} points,
\DynSignerRepoDiffFullSanLow{} to \DynSignerRepoDiffFullSanHigh{}).

Dropping the ref also loses information. Of the
\DynSignerBranchPooledPackages{} pooled packages that sign from several
branches, \DynSignerBranchPooledLine{} use branches named after a
maintenance line, so a project can mark per-line pipelines in the ref.
In sample~2, a pin that accepts
any tag and keeps branch refs cuts the full-SAN alerts from
\DynSignerFTwoPartAlerts{} to
\DynSignerFTwoTagWildcardTimestampAdjacentAlerts{} under adjacent memory
(Table~\ref{tab:dynamics-replication}). Verifiers can
therefore pin the repository and workflow file, accept any tag, and keep
branch refs for projects that sign from maintenance branches.

Key-id and publisher pins leave more residual alerts:
\DynResidualPublisherOrKeyAlerts{} of the \DynResidualAlerts{} residual
alerts in samples~1--3 mark a publisher or signing key new to the
package (Table~\ref{tab:dynamics-residual}). Apache projects' KEYS files
list the key in \DynDeclKeysPooledDeclaredAlerts{} of the
\DynDeclKeysPooledApacheAlerts{} new-key alerts on their packages
(Appendix~\ref{app:waterfall-orders}). A project's KEYS file can thus
put a new-key alert in context.

%% file: generated/tab_positive_controls.tex
\begin{table}[t]\centering\footnotesize
\caption{Real publishes with stolen credentials. No-downgrade flags the attacker in 5 of 12 incidents, exactly those whose package already carried provenance or attestations; identity pins flag almost none.}\label{tab:positive-controls}
\setlength{\tabcolsep}{2.6pt}
\begin{tabular*}{\columnwidth}{@{\extracolsep{\fill}}lllllllrr@{}}\toprule
 & & & \multicolumn{2}{c}{R1} & \multicolumn{2}{c}{R4} & \multicolumn{2}{c@{}}{Benign} \\\cmidrule(lr){4-5}\cmidrule(lr){6-7}\cmidrule(l){8-9}
Incident & Prior & pnpm & adj. & line & hist. & line & pnpm & line \\\midrule
eslint-scope (2018) & none & 0/1 & 0/1 & 0/1 & ? & ? & 0 & 0 \\
ua-parser-js (2021) & none & 0/3 & 0/3 & 0/3 & 0/3 & 0/3 & 0 & 0 \\
coa, rc (2021) & none & 0/9 & 0/9 & 0/9 & ? & ? & 0 & 0 \\
ultralytics$^p$ (2024) & att. & 2/2$^a$ & 1/2 & 2/2 & -- & -- & 0$^a$ & 0 \\
@rspack/core (2024) & prov. & 1/1$^{\ddagger}$ & 1/1 & 1/1 & ? & ? & 0 & 0 \\
eslint-config-prettier (2025) & prov. & 4/4$^{\ddagger}$ & 1/4 & 2/4 & 0/4 & 2/4 & 2 & 0 \\
nx$^{\dagger}$ (2025) & prov. & 8/8 & 1/8 & 8/8 & 0/8 & 0/8 & 0 & 0 \\
chalk, debug (2025) & none & 0/2 & 0/2 & 0/2 & 0/2 & 1/2 & 0 & 0 \\
@ctrl/tinycolor$^{\dagger}$ (2025) & none & 0/2 & 0/2 & 0/2 & 0/2 & 0/2 & 0 & 0 \\
posthog-node (2025) & none & 0/3 & 0/3 & 0/3 & ? & ? & 0 & 0 \\
litellm, telnyx$^p$ (2026) & none & 0/4$^a$ & 0/4 & 0/4 & -- & -- & 56$^a$ & 56 \\
axios (2026) & TP & 2/2 & 1/2 & 1/2 & 0/2 & 0/2 & 2 & 1 \\
\midrule
All (41 releases) & & 17 & 5 & 14 & 0 & 3 & 60 & 57 \\
\bottomrule
\end{tabular*}
\par\smallskip\parbox{\columnwidth}{\scriptsize\raggedright $k/n$: $k$ of the incident's $n$ malicious releases flagged. Prior: strongest earlier evidence (TP trusted publisher, prov.\ provenance, att.\ PyPI attestation). adj.: adjacent memory; line: per-line history; hist.: whole-package history. R2 flags none; R3 flags 1 (npm's 2025 registry-key rotation). Benign: flagged releases within $\pm$90 days of the incident. ultralytics: only 8.3.45 and 8.3.46, uploaded with an unrevoked API token, count; 8.3.41 and 8.3.42 came through the project's own workflow. $^p$PyPI. $^a$No pnpm on PyPI: the same rule on attestations. $^{\dagger}$Publisher inferred from other facts of the incident. $^{\ddagger}$Flag rests on an inferred absence of provenance. ?: publisher not recovered; --: PyPI names no per-release uploader. Benign flags for litellm, telnyx follow evidence adopted after the incident.}
\end{table}

%% file: generated/tab_declared_policies.tex
\begin{table}[t]\centering\footnotesize
\caption{pnpm's rule, its per-line variant, and two declared policies on the same releases.}\label{tab:declared-policies}
\setlength{\tabcolsep}{3pt}
\begin{tabular*}{\columnwidth}{@{\extracolsep{\fill}}lrrrrr@{}}\toprule
 & \multicolumn{2}{c}{Failed installs$^{a}$} & \shortstack[r]{Blocking\\releases} & \multicolumn{2}{c@{}}{\shortstack{Malicious\\refused}} \\\cmidrule(lr){2-3}\cmidrule(l){5-6}
Rule & Popular & \shortstack[r]{Recently\\updated} & refused & of 17 & of 24 \\\midrule
pnpm's rule (whole-package history) & 16 & 76 & 34 & 17 & 0 \\
Per-line history & 2 & 31 & 18 & 14 & 0 \\
Registered pipeline & 16 & 70/75 & 31/33 & 17 & 24/0 \\
Minimal per-line declaration & 2 & 37/43 & 24/27 & 14/15 & 0 \\
\bottomrule
\end{tabular*}
\par\smallskip\parbox{\columnwidth}{\scriptsize\raggedright Failed installs that remain, of the 16 popular and 76 recently updated installs that fail under pnpm's rule; blocking releases refused, of 34; malicious releases refused, of the 17 with and the 24 without earlier evidence. Two values give the two replay conventions: as replayed (a history keeps the releases the rule would have refused) and enforcing (a refused release never enters a later history, and the registered pipeline applies from the package's first release with evidence). $^{a}$At least: pnpm names one blocking release per failed install, so the installs on an admitted release are an upper bound on those a rule lets through.}
\end{table}

%% file: sections/case_studies.tex
\section{Case Studies}\label{sec:case-studies}

\paragraph*{\hspace{-\parindent}\bf semver 6.3.1: a security fix reported as a possible takeover.}
\texttt{semver}, downloaded \DynCaseSemverWeeklyDownloads{} times a
week, lies outside our samples; its \texttt{6.3.1} release causes
\DynEcoSecurityBackportOneTargets{} of the \DynEcoTotalFailing{} failed
installs of Finding~1. In June 2023, the project fixed a regular-expression
denial-of-service vulnerability (\DynCaseSemverAdvisoryCve{}) in
\texttt{7.5.2}, published with provenance. In July, a
maintainer backported the fix to the 6.x line as \texttt{6.3.1} and
published it without provenance. pnpm orders history by publish time, so
the earlier 7.x provenance sets the rank required of \texttt{6.3.1}, and
pnpm~\DynPnpmVersion{} refuses the fix as a ``possible package
takeover''~\cite{pnpm_issue_10202}. No 6.x release has followed, so every
package whose dependency range resolves to \texttt{6.3.1} inherits the
block, as our run of \texttt{@babel/core} \texttt{7.29.7}, the newest 7.x
of the Babel compiler, shows.

\paragraph*{\hspace{-\parindent}\bf uuid: one fix, three lines, two verdicts.} \texttt{uuid}
published \texttt{14.0.0} through a trusted publisher on April~19, 2026.
On April~29, the project backported a fix for
\DynCaseUuidAdvisoryCve{} to three older lines: \texttt{13.0.1} manually,
then \texttt{12.0.1} and \texttt{11.1.1} through the trusted publisher. All
three follow \texttt{14.0.0} in publish time, so pnpm~\DynPnpmVersion{}
compares each with it and blocks only \texttt{13.0.1}
(Table~\ref{tab:dynamics-pnpm-blocks}). The same fix arriving through CI
minutes later gives a reviewer context that the drop in evidence alone
cannot supply. On May~4, \texttt{uuid} re-released it through the
trusted publisher as \texttt{13.0.2}, which pnpm passes.

%% file: sections/discussion.tex
\section{Discussion}\label{sec:discussion}

\paragraph*{\hspace{-\parindent}\bf Consumer tools.} Consumers should set the cooldown
explicitly, which makes it strict (\S\ref{sec:rules}), and keep the age
cutoff longer than the cooldown; \DynAdoptionIgnoreAfterLeCooldown{}
sampled GitHub repositories set a cutoff no longer than the cooldown
(Finding~4). An exemption should name an exact version, which admits one
reviewed backport while every other release stays checked. Exempting
all of \texttt{axios} after pnpm blocks the maintainers' \texttt{0.30.3}
also admits the attacker's \texttt{0.30.4}, yet
\DynAdoptionHeadlineReposWhole{} repositories exempt whole packages, and
pnpm removes unused exemptions only on request (\S\ref{sec:rules}). Tools
should present alerts as prompts for review, with context beyond registry
evidence (\S\ref{sec:case-studies}).

\paragraph*{\hspace{-\parindent}\bf Maintainers.} pnpm compares a stable
release with every earlier stable release, on any line, so a patch for an
older line must meet the highest rank, as
\texttt{uuid}'s \texttt{13.0.2} shows (\S\ref{sec:case-studies}). For
\texttt{semver}, that rank is now a trusted publisher, so a
\texttt{6.3.2} with provenance alone would be blocked. Since
pnpm~\DynStagedRankSince{}, a package that has staged one release must
stage every later one, including CI releases (Finding~5).

\paragraph*{\hspace{-\parindent}\bf The Web precedent.} HSTS lets a site
require HTTPS for a set period~\cite{rfc6797}, and HPKP let a site pin its
keys~\cite{rfc7469}. Both read a declaration from the site, whereas
pnpm's policy and identity pins infer the required rank and identities
from past releases. Measurements found misconfigured HSTS and sparse, often incorrect
HPKP~\cite{kranch2015pinning,amann2017diginotar}; HPKP could lock users
out after a key change, and Chrome removed it~\cite{palmer2017hpkp,chromestatus2018hpkp}.
The Web
also uses declarations and public logs: every public
certificate authority must check a domain's CAA record before
issuing~\cite{rfc8659,cabforum2017ballot187,scheitle2018caa}, and Chrome
requires Certificate Transparency for publicly trusted TLS
certificates~\cite{chrome2026ctpolicy,scheitle2018ct}.

\paragraph*{\hspace{-\parindent}\bf Registries.} Registries can let a package
declare which pipelines may publish it: each
trusted publisher's repository and workflow, whether maintainer accounts
may still upload, and which channels and lines each pipeline serves. The
registry then refuses other uploads at publication, and consumers can read
the declaration, which carries the trade-off of Finding~3. A project that
declares only its pipeline refuses the malicious releases that pnpm flags
and must publish its backports through that pipeline, as \texttt{uuid}
did. A project that keeps a
manual line keeps that line's exposure: as replayed, the \texttt{0.30}
line of \texttt{axios} is manual and admits \texttt{axios}~\texttt{0.30.4}.
Enforced, the line never becomes manual, so the registry refuses every
release on it without evidence, including the maintainers'.

Parts of such declarations are deployed. npm, PyPI, and crates.io
registrations name the repository and workflow allowed to
publish~\cite{npm_trusted_publishers,pypi_trusted_publishers,crates2026trusted},
npm's staged publishing adds a second-factor
approval~\cite{npm_staged_publishing}, and npm and crates.io let a package
refuse token uploads~\cite{npm_trusted_publishers,crates2026trustpubonly}.
Of the \DynDeclCratesAllWithTrustedPublishing{} sampled crates with a
trusted publisher, \DynDeclCratesAllTrustpubOnlyOfTrustedPublishing{} do;
after adopting trusted publishing, these
\DynDeclCratesAfterAdoptionCrates{} published
\DynDeclCratesAfterAdoptionTokenReleases{} of
\DynDeclCratesAfterAdoptionReleases{} releases with a token. On the
reconstructed incidents, such a setting refuses the
\DynDeclReplayIncidentNoTokenRefuse{} malicious releases published with a
stolen token alone and none of the
\DynDeclReplayIncidentAccountReleases{} published from a taken-over
account. Registries know whether a token, an interactive session, or a
trusted publisher authenticated each upload, but public records do not
separate the first two. We recommend publishing this field, which
separates an interactive backport from a stolen-token release.

An attacker who holds the account can also edit a declaration; delaying
such changes and recording them in a public event
log~\cite{openssf2024principles} would give consumers a window to reject
releases under a new policy. A declaration should also bind evidence to
the registered repository, which pnpm's ranks do not test; a
compromised workflow passes any of these policies (Finding~2). Our
proposal places trust in the registry; PEP~480 limits that trust to keys
kept offline~\cite{pep480}.

\paragraph*{\hspace{-\parindent}\bf Limitations.} Our counts describe the sampled packages and two
npm populations at the observation dates and give no registry-wide
rates. The recently updated population covers two days of publishing,
mostly by scoped packages; the direction of bias in its failure rate is
unknown. The main failure rates apply to pnpm~\DynPnpmVersion{},
\DynPnpmMidVersion{}, and \DynPnpmCurrentVersion{} with their default
settings and exact requests for each package's latest version. They apply
only to consumers who enable the opt-in policy. The sample, incident, and
cutoff results use the ranking of pnpm~\DynPnpmVersion{}. Code search
excludes private repositories, so the adoption count is a lower bound.

Our replays assume that malicious releases carry no evidence
(Finding~2) and that attackers add none; both assumptions may overstate
how many attacks the policy flags; attackers who add evidence pass most
flagged releases (Finding~2). Removal times are partly
upper bounds, and advisories often follow a malicious release's removal;
both can overstate what a cutoff leaves unchecked. Declared
policies are inferred from package histories; what projects would
declare is open. The census misses
incidents without a malware advisory, such as those of
\texttt{@rspack/core} and \texttt{posthog-node}.

Nearly all keyless identities come from GitHub Actions, so the effect of
the ref reflects GitHub's certificate format. We read identities from
served certificates without verifying signatures. Our histories end early
in the adoption of trusted publishing~\cite{githubchangelog2025npmtrusted};
wider adoption gives the policy more earlier evidence for blocks and
flags.

\paragraph*{\hspace{-\parindent}\bf Ethics and disclosure.} All data come
from public endpoints through read-only requests, at most
\DynEthicsWorkersMax{} at a time per collector, and the pnpm runs executed
no package code. We report public configuration files in aggregate,
archive them without e-mail addresses, and compare npm account e-mail
addresses only as hashes. We reported our pnpm findings privately to its
maintainers and notified the maintainers of blocked packages.

%% file: sections/related_work.tex
\section{Related Work}\label{sec:related-work}

\paragraph*{\hspace{-\parindent}\bf Measurements of signing and provenance.} Schorlemmer et al.\
measure signing quantity, quality, and longitudinal trends across four
registries~\cite{schorlemmer2024registrysigning}. Kalu et al.\ study adoption
and usability through industry interviews and a Sigstore case
study~\cite{kalu2025industrysigning,kalu2026usability}. We measure what
a deployed rule and identity pins decide from this evidence.

\paragraph*{\hspace{-\parindent}\bf Publication authorization and verification.} TUF lets a
package repository delegate packages to signing keys~\cite{tufspec2026};
in-toto layouts name who may perform each
step~\cite{torresarias2019intoto}. Diplomat and PEPs~458 and~480 adapt TUF
to community repositories~\cite{kuppusamy2016diplomat,pep458,pep480}.
Speranza serves per-package authorized
signers~\cite{merrill2023speranza}, and DiVerify adds signing
conditions~\cite{okafor2024diverify}. SLSA
names three ways to form expectations about provenance: trust on first use, by the producer,
and in source~\cite{slsa2025v12}. We measure the first and approximate the second
(Finding~3).

\paragraph*{\hspace{-\parindent}\bf Attacks and update analysis.} Prior work studies
attacks on package managers~\cite{cappos2008mirror} and
registries~\cite{gu2023registries} and measures maintainer reach~\cite{zimmermann2019}, takeover
paths~\cite{zahan2022weaklinks}, and malicious packages and their
detectors~\cite{ohm2020backstabbers,ladisa2023sok,duan2021measuring,guo2023malicious,guo2026npmdetectors}.
RogueOne compares the data flows of successive
versions~\cite{sofaer2024rogueone}; our rules compare publication
evidence.

%% file: sections/conclusion.tex
\section{Conclusion}\label{sec:conclusion}

Consumer rules that compare publication evidence across releases trade
blocked maintenance releases against missed attacks. With
pnpm~\DynPnpmVersion{}'s defaults, the opt-in no-downgrade policy breaks
installs of \DynEcoTopFailShare{} of the popular npm packages tested.
All are JavaScript build, lint, or test tools, and most of these
failures trace to one security backport that users exempt by hand. The
policy can flag attacks only where earlier releases carried stronger
evidence, as for \DynCensusAllPriorEvidenceShare{} of attacked packages
in our census
(\DynCensusIncidentBootPriorEvidenceLo{}--\DynCensusIncidentBootPriorEvidenceHi{}
when incidents are resampled), and a strict one-day cooldown already
keeps out most reconstructed malicious releases. Comparing within each
maintenance line admits backports and attacks alike, and a registry
declaration makes that choice explicit. In our replay, declaring lines
manual removes at most \DynDeclReplayAsReplayedEcoPerLineFailuresRemovedBoth{} of the
\DynDeclReplayEcoFailuresBoth{} failed installs and admits
\DynDeclReplayEnforcingIncidentPerLineAdmitWithEarlierEvidence{}
or~\DynDeclReplayAsReplayedIncidentPerLineAdmitWithEarlierEvidence{} of
the \DynDeclReplayIncidentWithEarlierEvidence{} malicious releases that
pnpm flags. Registries can also publish how each release was
authenticated.

%% file: sections/appendix.tex
% Tables are numbered in the order of the sections that discuss them. Each
% table is input before the text of the page on which it should appear, so
% the two-column tables reach the top of the next page.
\section{Samples}\label{app:samples}

Table~\ref{tab:dynamics-samples} lists the packages of the four samples by
registry and stratum (\S\ref{sec:samples}). It counts the packages that
we collected and analyzed. In samples~2--4, a cell falls
short of its target when too few eligible candidates remain after earlier
samples or when the registry no longer serves a selected package.

\section{Alert Decomposition}\label{app:waterfall-orders}
The decomposition of \S\ref{sec:attribution} relaxes each rule's naive
variant with five mechanisms. \emph{Registry} ignores changes caused by a
registry-wide event, here npm's key rotation; \emph{channel} and
\emph{line} compare a release only with earlier releases on its channel or
maintenance line; \emph{recurrence} ignores values and evidence drops that
the package has shown before; and \emph{reference} removes the ref from the
certificate identity. The mechanisms overlap, so
Table~\ref{tab:dynamics-waterfall-orders} reports the alerts attributed to
each mechanism as a mean and range over all 120 orderings; the residual
alerts are the same in every ordering. In every sample and every ordering,
the reference removes at least \DynDecompFOneRTwoRefMinShare{} of the
naive signer-pin alerts.
Table~\ref{tab:dynamics-residual} names the residual alerts. Apache KEYS
files list the key in \DynDeclKeysPooledDeclaredAlerts{} of the
\DynDeclKeysPooledApacheAlerts{} new-key alerts on packages of Apache
projects. A signature names only its key, so the
\DynDeclKeysPooledOtherAlerts{} new-key alerts elsewhere stay
unclassified.

\input{generated/tab_app_samples}
\input{generated/tab_app_waterfall_orders}
\input{generated/tab_residual}
\input{generated/tab_pnpm_blocks}

\section{pnpm: History Scope, Age Cutoff, and Ranking}\label{app:pnpm-blocks}
\input{generated/tab_app_history_scope}
\input{generated/tab_ignore_after_no_current}
\input{generated/tab_pnpm_transition}
Table~\ref{tab:dynamics-pnpm-blocks} lists every stable release that pnpm's
rule blocks in samples~1--3 (Finding~3).
Under the history scopes of Table~\ref{tab:dynamics-history-scope}, PyPI
releases that dropped attestations raise
\DynRegROneStableBlocksPypiFTwo{} alerts in sample~2 and
\DynRegROneStableBlocksPypiFThree{} in sample~3, and crates.io adds
\DynRegROneStableBlocksCratesFTwo{} in sample~2. The alerts that remain
under a per-line history concern drops within a line, and none of the
tested scopes passes these releases.
Table~\ref{tab:ignore-after} reports every age cutoff we ran (Finding~4);
its Incidents columns also give the malicious releases that a strict
cooldown of the same length keeps out (\S\ref{sec:rq1}).
Table~\ref{tab:pnpm-transition} pairs each install's outcome under
pnpm~\DynPnpmCurrentLegacyVersion{} and \DynPnpmCurrentVersion{}
(Finding~5). Of the \DynPnpmCurrentLegacyRandomPassingNowFailing{} installs
that only the later release fails,
\DynPnpmCurrentNowFailingCauseStaged{} follow a staged publish:
\DynPnpmCurrentNowFailingStagedWithTrustedPublisher{} name a release with
a trusted publisher and provenance and
\DynPnpmCurrentNowFailingStagedWithoutEvidence{} one without evidence.
The other \DynPnpmCurrentNowFailingCauseTpNoProvenance{} name a
trusted-publisher release without provenance.

\section{Declared Policies on Sampled Releases}\label{app:declared-samples}
Among the \DynDeclReplaySampleStableNpmReleases{} stable npm releases of
samples~1--4, pnpm's rule blocks
\DynDeclReplayAsReplayedSampleStableNpmDeployedRefuse{}. Of their
\DynDeclReplaySampleStableNpmPackages{} packages,
\DynDeclReplaySampleStableNpmPackagesNoPipeline{} have no release with
evidence and thus no pipeline to register. As replayed, the
registered-pipeline policy refuses
\DynDeclReplayAsReplayedSampleStableNpmRegisteredRefuse{} releases.
Enforced from each package's first release with evidence, it refuses
\DynDeclReplayEnforcingSampleStableNpmRegisteredRefuse{}; pnpm's rule
passes \DynDeclReplayEnforcingSampleStableNpmRegisteredRefusePassesDeployed{}
of them, whose evidence names another repository. The per-line
declaration refuses \DynDeclReplayAsReplayedSampleStableNpmPerLineRefuse{}
releases as replayed and
\DynDeclReplayEnforcingSampleStableNpmPerLineRefuse{} when enforced, and
it admits \DynDeclReplayAsReplayedSampleStableNpmPerLineAdmitDeployedRefuses{}
and \DynDeclReplayEnforcingSampleStableNpmPerLineAdmitDeployedRefuses{} of
the releases that pnpm's rule blocks. On the
\DynDeclReplaySampleStablePypiReleases{} stable PyPI releases, whose
histories start with the samples, R1 alerts on
\DynDeclReplayAsReplayedSampleStablePypiDeployedRefuse{}; the registered
pipeline refuses \DynDeclReplayAsReplayedSampleStablePypiRegisteredRefuse{} as
replayed and \DynDeclReplayEnforcingSampleStablePypiRegisteredRefuse{}
when enforced, and the per-line declaration
\DynDeclReplayAsReplayedSampleStablePypiPerLineRefuse{} and
\DynDeclReplayEnforcingSampleStablePypiPerLineRefuse{}.

\section{Replication}\label{app:replication}
\input{generated/tab_app_replication}
\input{generated/tab_signer_pins}
Table~\ref{tab:dynamics-replication} lists the values behind Findings~6
and~7 for each sample, including sample~4, which shares no repository with
the others.

% Every appendix float must be placed before this text ends: acmart
% balances the last page (\AtEndDocument{\balance}), and balance.sty drops
% one-column floats that are still pending there. The Replication text
% above reaches the last page, so its two tables are placed with it.

%% file: generated/tab_app_samples.tex
\begin{table}[!t]\centering\footnotesize
\caption{Packages per registry and stratum. Sample~1 holds 251 releases per package against 36, 43, and 34 in samples~2--4, because its activity strata favor frequent publishers. Samples~2--4 target 30, 40, and 50 packages per stratum; every package has at least four releases between April~6, 2024 and September~2, 2026. In sample~4, the popular pools of crates.io and RubyGems hold no unused eligible package.}\label{tab:dynamics-samples}
\setlength{\tabcolsep}{2.8pt}
\begin{tabular}{@{}lrrrrrrrrr@{}}\toprule
 & \multicolumn{3}{c}{S1} & \multicolumn{2}{c}{S2} & \multicolumn{2}{c}{S3} & \multicolumn{2}{c}{S4} \\
\cmidrule(l{2pt}r{2pt}){2-4}\cmidrule(l{2pt}r{2pt}){5-6}\cmidrule(l{2pt}r{2pt}){7-8}\cmidrule(l{2pt}r{2pt}){9-10}
Registry & High & Med. & Low & Pop. & Rand. & Pop. & Rand. & Pop. & Rand. \\\midrule
crates.io & 8 & 9 & 8 & 30 & 28 & 33 & 40 & -- & 50 \\
Maven & 10 & 10 & 10 & 29 & 30 & 40 & 40 & 49 & 49 \\
npm & 10 & 10 & 10 & 30 & 28 & 40 & 38 & 41 & 49 \\
PyPI & 10 & 10 & 10 & 30 & 30 & 40 & 38 & 50 & 49 \\
RubyGems & 10 & 10 & 10 & 30 & 30 & 30 & 36 & -- & 29 \\
\midrule
Packages & \multicolumn{3}{c}{145} & \multicolumn{2}{c}{295} & \multicolumn{2}{c}{375} & \multicolumn{2}{c}{366} \\
Releases & \multicolumn{3}{c}{36{,}465} & \multicolumn{2}{c}{10{,}586} & \multicolumn{2}{c}{16{,}071} & \multicolumn{2}{c}{12{,}471} \\
Releases per package & \multicolumn{3}{c}{251} & \multicolumn{2}{c}{36} & \multicolumn{2}{c}{43} & \multicolumn{2}{c}{34} \\
\bottomrule\end{tabular}
\end{table}

%% file: generated/tab_app_waterfall_orders.tex
\begin{table}[tbp]\centering\footnotesize
\caption{Alerts each mechanism removes in samples~1--3: mean and range over all 120 orderings of the five mechanisms. The ordering never changes the residual. In every ordering, the reference removes more than half of the naive signer-pin alerts. Mechanisms that remove no alert of a rule are omitted.}\label{tab:dynamics-waterfall-orders}
\setlength{\tabcolsep}{4pt}
\begin{tabular}{@{}lrrrrrr@{}}\toprule
 & \multicolumn{2}{c}{S1} & \multicolumn{2}{c}{S2} & \multicolumn{2}{c}{S3} \\
\cmidrule(l{2pt}r{2pt}){2-3}\cmidrule(l{2pt}r{2pt}){4-5}\cmidrule(l{2pt}r{2pt}){6-7}
Step & Mean & Range & Mean & Range & Mean & Range \\\midrule
\multicolumn{7}{@{}l}{\emph{R1 attestation downgrade}} \\
\quad naive alerts & 5 &  & 190 &  & 160 &  \\
\quad $-$ channel & 0.0 & 0 & 90.8 & 2--180 & 75.0 & 0--150 \\
\quad $-$ line & 0.0 & 0 & 1.2 & 1--2 & 0.0 & 0 \\
\quad $-$ recurrence & 2.0 & 2 & 91.0 & 2--180 & 75.0 & 0--150 \\
\quad residual & 3 & (constant) & 7 & (constant) & 10 & (constant) \\
\addlinespace[2pt]
\multicolumn{7}{@{}l}{\emph{R2 keyless signer pin}} \\
\quad naive alerts & 3{,}544 &  & 690 &  & 642 &  \\
\quad $-$ channel & 56.5 & 0--120 & 2.0 & 0--11 & 7.9 & 0--28 \\
\quad $-$ line & 1.2 & 0--4 & 64.4 & 0--201 & 55.0 & 0--169 \\
\quad $-$ recurrence & 771.7 & 0--1{,}541 & 73.8 & 8--204 & 64.2 & 0--188 \\
\quad $-$ reference & 2{,}710.6 & 1{,}999--3{,}428 & 543.8 & 475--676 & 505.0 & 440--631 \\
\quad residual & 4 & (constant) & 6 & (constant) & 10 & (constant) \\
\addlinespace[2pt]
\multicolumn{7}{@{}l}{\emph{R3 key-id pin}} \\
\quad naive alerts & 43 &  & 47 &  & 125 &  \\
\quad $-$ registry & 9.0 & 9 & 27.0 & 27 & 46.0 & 46 \\
\quad $-$ channel & 0.7 & 0--2 & 0.0 & 0 & 3.0 & 0--8 \\
\quad $-$ line & 9.0 & 0--20 & 0.0 & 0 & 5.7 & 0--14 \\
\quad $-$ recurrence & 15.3 & 5--25 & 3.0 & 3 & 37.3 & 31--46 \\
\quad residual & 9 & (constant) & 17 & (constant) & 33 & (constant) \\
\addlinespace[2pt]
\multicolumn{7}{@{}l}{\emph{R4 publisher pin}} \\
\quad naive alerts & 211 &  & 115 &  & 164 &  \\
\quad $-$ channel & 21.3 & 0--44 & 1.0 & 0--2 & 2.5 & 0--5 \\
\quad $-$ line & 29.8 & 0--61 & 1.0 & 0--2 & 1.0 & 0--2 \\
\quad $-$ recurrence & 110.8 & 61--162 & 49.0 & 47--51 & 51.5 & 48--55 \\
\quad residual & 49 & (constant) & 64 & (constant) & 109 & (constant) \\
\bottomrule\end{tabular}
\end{table}

%% file: generated/tab_residual.tex
\begin{table*}[t]\centering\footnotesize
\caption{Named residual after decomposition, pooled over samples~1--3. Most alerts that no mechanism removes mark a new publisher or a Maven signing key the package has not used before. Rows cover all 321 residual alerts; the 10 repository moves are grouped by kind.}\label{tab:dynamics-residual}
\setlength{\tabcolsep}{4pt}
\begin{tabular}{@{}lllcr@{}}\toprule
Class & Before $\to$ after & Example release & Samples & Alerts \\\midrule
New publisher & account $\to$ another account & crates.io: google-cloud-retail-v2@0.2.0 & 1, 2, 3 & 131 \\
 & account/CI $\to$ trusted publisher (or reverse) & npm: @octokit/openapi-types-ghes-3.18@29.0.0 & 1, 2, 3 & 81 \\
 & CI $\to$ another CI identity & PyPI: muxu-io-mqtt-connector@1.0.0 & 1, 3 & 5 \\
\addlinespace[2pt]New key & Maven signing key not used before & Maven: org.apache.lucene:lucene-core@10.0.0 & 1, 2, 3 & 57 \\
\addlinespace[2pt]Repository move & new owner: rustwasm/wasm-bindgen $\to$ wasm-bindgen/wasm-bindgen, and 3 others & crates.io: wasm-bindgen-macro-support@0.2.101 & 1, 2, 3 & 8 \\
 & same owner: eslint-scope, espree $\to$ eslint/js, and 2 others & npm: eslint-scope@8.1.0 & 2, 3 & 5 \\
 & \emph{repo} $\to$ \emph{repo}-release: matplotlib, scikit-learn & PyPI: scikit-learn@1.9.1 & 2, 3 & 4 \\
\addlinespace[2pt]Other evidence & attestation dropped, same or no publisher & npm: undici-types@6.19.3 & 1, 2, 3 & 17 \\
\addlinespace[2pt]New workflow & new workflow file, same signer repository & npm: @graphql-hive/gateway-testing@6.0.6 & 1, 2, 3 & 13 \\
\bottomrule\end{tabular}
\par\smallskip\parbox{\textwidth}{\scriptsize\raggedright R4 compares the publisher that the registry names. crates.io names no account for a trusted-publisher upload, so R4 evaluates none of the 343 crates.io releases with a trusted-publishing record in these samples. Of the 81 alerts in the trusted-publisher row, 68 are npm packages that move from an account to a trusted publisher.}
\end{table*}

%% file: generated/tab_pnpm_blocks.tex
\begin{table*}[t]\centering\footnotesize
\caption{Stable releases that pnpm's no-downgrade check blocks in samples~1--3; sample~1 has none. One evidence lapse blocks every later release without the evidence, across major lines, until a release restores it.}\label{tab:dynamics-pnpm-blocks}
\setlength{\tabcolsep}{4.5pt}
\begin{tabular}{@{}lllllrl@{}}\toprule
Package & Sample & Blocked versions & Published & Evidence & Releases & Next pass \\\midrule
\texttt{@landstrip/landstrip-darwin-arm64} & S3 & 0.17.27--0.19.4 & Jul 19--Sep 26, 2026 & provenance$\to$none & 66 & none in capture \\
\texttt{undici-types} & S2 & 6.19.3--7.0.0 & Jul 22--Nov 28, 2024 & provenance$\to$none & 8 & 7.1.0 \\
\texttt{undici-types} & S2 & 7.2.0 & Dec 18, 2024 & provenance$\to$none & 1 & 7.2.1 \\
\texttt{ts-api-utils} & S2 & 2.1.1--2.3.0 & Dec 27--28, 2025 & provenance$\to$none & 3 & 2.4.0 \\
\texttt{chokidar} & S3 & 4.0.2--4.0.3 & Dec 16--18, 2024 & provenance$\to$none & 2 & 5.0.0 \\
\texttt{uuid} & S2 & 13.0.1$^\dagger$ & Apr 29, 2026 & trusted publisher$\to$none & 1 & 12.0.1 (13.x: 13.0.2) \\
\texttt{@octokit/openapi-types-ghes-3.18} & S3 & 29.0.0 & Aug 27, 2026 & trusted publisher$\to$none & 1 & 29.0.1 \\
\midrule
Total & & & & & 82 & \\
\bottomrule
\multicolumn{7}{@{}l@{}}{\rule{0pt}{2.2ex}Each row is one run of consecutive blocked stable releases in publish order. Next pass: the next stable release by publish time that pnpm passes;} \\
\multicolumn{7}{@{}l@{}}{for a backport, the first passing release on its own line follows in parentheses.} \\
\multicolumn{7}{@{}l@{}}{$^\dagger$Security backport, the fix for CVE-2026-41907 (GHSA-w5hq-g745-h8pq).}
\end{tabular}
\end{table*}

%% file: generated/tab_app_history_scope.tex
\begin{table}[tbp]\centering\footnotesize
\caption{Stable releases that R1 with history memory flags in samples~1--3 (npm, PyPI, crates.io), by history scope. A per-channel history removes only typescript-eslint alerts; a per-line history removes 3 more in sample~2 and 63 in sample~3; lapses within a line remain. Of the 13 and 70 npm releases in the per-channel row, pnpm blocks 13 and 69 (Table~\ref{tab:dynamics-pnpm-blocks}); R1 and pnpm rank the evidence of \texttt{whatwg-url} 17.1.1 differently. Rel.: releases; Pkgs.: packages.}\label{tab:dynamics-history-scope}
\setlength{\tabcolsep}{4pt}
\begin{tabular}{@{}lrrrrrr@{}}\toprule
 & \multicolumn{2}{c}{S1} & \multicolumn{2}{c}{S2} & \multicolumn{2}{c}{S3} \\
\cmidrule(l{2pt}r{2pt}){2-3}\cmidrule(l{2pt}r{2pt}){4-5}\cmidrule(l{2pt}r{2pt}){6-7}
History scope & Rel. & Pkgs. & Rel. & Pkgs. & Rel. & Pkgs. \\\midrule
Whole package (all releases) & 0 & 0 & 174 & 8 & 288 & 10 \\
Per channel (pnpm $\geq$10.24) & 0 & 0 & 22 & 6 & 136 & 8 \\
Per maintenance line & 0 & 0 & 19 & 5 & 73 & 7 \\
\bottomrule\end{tabular}
\end{table}

%% file: generated/tab_ignore_after_no_current.tex
\begin{table}[t]\centering\footnotesize
\caption{With \texttt{trustPolicyIgnoreAfter}, pnpm skips its no-downgrade check for releases older than the cutoff. Each row counts the installs that still fail and the malicious releases that are still checked because their whole live window is shorter than the cutoff.}\label{tab:ignore-after}
\setlength{\tabcolsep}{3.6pt}
\begin{tabular}{@{}lrrrrrrr@{}}\toprule
 & \multicolumn{2}{c}{Failing installs} & \multicolumn{3}{c}{Incidents} & \multicolumn{2}{c@{}}{Census} \\\cmidrule(lr){2-3}\cmidrule(lr){4-6}\cmidrule(l){7-8}
Cutoff & Popular & \shortstack[r]{Recently\\updated} & All & Flagged$^{a}$ & Inc. & All & Flagged \\\midrule
none (default) & 16 & 77 & 41 & 17 & 12 & 100\% & 100\% \\
1 day & 0 & n/m & 36 & 12 & 10 & 76\% & 45\% \\
7 days & 0 & 12 & 40 & 16 & 11 & 83\% & 48\% \\
30 days & 0 & 19 & 40 & 16 & 11 & 90\% & 78\% \\
365 days & 0 & 35 & 40 & 16 & 11 & $>$99\% & 100\% \\
\bottomrule
\end{tabular}
\par\smallskip\parbox{\columnwidth}{\scriptsize\raggedright Failing installs: live pnpm 10.29.3 runs on 2026-09-29, same 1{,}000 popular and 997 recently updated packages. n/m (not measurable): the youngest release blocking a recently updated install was 2.7 days old then, so no install could fail under a shorter cutoff and the cell would be 0 by construction. Incidents: 41 malicious releases, 12 incidents (Inc.: all releases checked); removal times: 17 exact from a source, 20 upper bounds, 3 inferred, 1 unknown (unchecked). Census: share of 4{,}229 malicious versions with an advisory within the cutoff (live-window proxy). Flagged: blocked with no cutoff (17 and 507); $^{a}$with 2 PyPI releases scored by R1 plus the cutoff (pnpm runs only on npm). The Incidents columns also give the counts for a strict release-age cooldown of the row's length: a release that is still checked was removed before it reached that age, so such a cooldown never installs it.}
\end{table}

%% file: generated/tab_pnpm_transition.tex
\begin{table}[t]\centering\footnotesize
\caption{Outcome of each install under pnpm 10.29.3 and under pnpm 12.9.1, both with \texttt{trustPolicy: no-downgrade} and no cooldown, on the same day. Rows pair the two outcomes; cells count packages.}\label{tab:pnpm-transition}
\setlength{\tabcolsep}{4pt}
\begin{tabular*}{\columnwidth}{@{\extracolsep{\fill}}lrr@{}}\toprule
pnpm 10.29.3 $\to$ pnpm 12.9.1 & Popular & \shortstack[r]{Recently\\updated} \\\midrule
Fails the check in both & 15 & 51 \\
Fails the check $\to$ installs, older version resolved & 1 & 12 \\
Fails the check $\to$ installs as without the policy & 0 & 5 \\
Fails the check $\to$ fails for another reason & 0 & 8 \\
\addlinespace[2pt]
Installs $\to$ fails the check & 0 & 7 \\
Installs in both & 984 & 895 \\
\addlinespace[2pt]
Fails for another reason in both & 0 & 17 \\
Fails for another reason $\to$ installs & 0 & 2 \\
\midrule
Packages & 1{,}000 & 997 \\
\bottomrule
\end{tabular*}
\par\smallskip\parbox{\columnwidth}{\scriptsize\raggedright Left of the arrow: pnpm 10.29.3; right: pnpm 12.9.1; one row without an arrow per unchanged outcome. Runs of 2026-10-04, \texttt{minimumReleaseAge: 0} for both clients, each package counted once. ``Older version resolved'': the newest version in a dependency's range fails the check and the client installs an older one that passes it. ``Installs as without the policy'': the lockfile equals that of the same client without the policy. Classes with no install are omitted.}
\end{table}

%% file: generated/tab_app_replication.tex
\begin{table}[tbp]\centering\footnotesize
\caption{Descriptive values in all four samples; sample~4 (S4) shares no repository with samples~1--3. The per-release ref in the certificate identity multiplies signer-pin alerts \mbox{29--58$\times$} in samples~1--3 and 5.1$\times$ in sample~4; the R1 predecessor ratio falls to 1.2$\times$ in sample~4. ``--'': not defined (sample~1 has no popular or random stratum, or no stable-only alert).}\label{tab:dynamics-replication}
\setlength{\tabcolsep}{4pt}
\begin{tabular}{@{}lrrrr@{}}\toprule
Metric & S1 & S2 & S3 & S4 \\\midrule
\multicolumn{5}{@{}l}{\emph{R2 keyless signer pin}} \\
\quad Alerts, full SAN / SAN w/o ref & 29.0$\times$ & 49.3$\times$ & 58.4$\times$ & 5.1$\times$ \\
\quad Share removed by the ref & 76.5\% & 78.8\% & 78.7\% & 76.7\% \\
\quad Alerting packages silent at repository level & 95.2\% & 90.6\% & 91.3\% & 87.2\% \\
\quad Share of alerts on top-3 packages & 75.1\% & 40.1\% & 38.3\% & 45.5\% \\
\quad npm packages alerting, full SAN $\to$ w/o ref & 9$\to$1 & 18$\to$5 & 20$\to$7 & 15$\to$4 \\
\quad PyPI packages alerting, full SAN $\to$ w/o ref & 12$\to$3 & 14$\to$1 & 26$\to$2 & 32$\to$4 \\
\quad Alerts, tags wildcarded (adjacent memory) & 3{,}009 & 232 & 228 & 228 \\
\quad Alerts, tags wildcarded (history memory) & 1{,}466 & 26 & 38 & 55 \\
\quad Packages alerting, tags wildcarded (adjacent) & 10 & 14 & 21 & 15 \\
\quad Packages alerting, tags wildcarded (history) & 10 & 14 & 22 & 15 \\
\addlinespace[2pt]
\multicolumn{5}{@{}l}{\emph{R1 attestation downgrade}} \\
\quad Alerts, timestamp / stable-only & -- & 27.1$\times$ & 20.0$\times$ & 1.2$\times$ \\
\quad Packages with a stable downgrade & 0 & 6 & 8 & 5 \\
\quad \ldots{} regaining the prior rank & 0.0\% & 83.3\% & 50.0\% & 60.0\% \\
\addlinespace[2pt]
\multicolumn{5}{@{}l}{\emph{Provenance on stable releases (packages)}} \\
\quad Adopt, never drop & 17 & 32 & 49 & 48 \\
\quad Drop once & 0 & 5 & 8 & 5 \\
\quad Drop more than once & 0 & 1 & 0 & 0 \\
\addlinespace[2pt]
\multicolumn{5}{@{}l}{\emph{Keyless-signing share of packages}} \\
\quad npm, popular & -- & 60.0\% & 57.5\% & 39.0\% \\
\quad npm, random & -- & 25.9\% & 23.7\% & 27.7\% \\
\quad PyPI, popular & -- & 40.0\% & 52.5\% & 50.0\% \\
\quad PyPI, random & -- & 26.7\% & 31.6\% & 20.4\% \\
\bottomrule\end{tabular}
\end{table}

%% file: generated/tab_signer_pins.tex
\begin{table}[t]\centering\footnotesize
\caption{Packages that alert under a signer pin (R2), by the identity it compares. Under the full identity, 74.6\%--76.7\% of the packages with at least two keyless-signed releases alert in every sample; without the Git ref, 12.7\%--15.0\% do.}\label{tab:signer-pins}
\setlength{\tabcolsep}{3pt}
\begin{tabular*}{\columnwidth}{@{\extracolsep{\fill}}lrrrrrr@{}}\toprule
 & & & & & \multicolumn{2}{c@{}}{S2--S4 pooled} \\\cmidrule(l){6-7}
Identity compared & S1 & S2 & S3 & S4 & Share & 95\% CI \\\midrule
Full SAN & 21 & 32 & 46 & 47 & 75.8\% & 68.7\%--81.7\% \\
Tag refs wildcarded & 10 & 14 & 21 & 15 & 30.3\% & 23.8\%--37.7\% \\
SAN without ref & 4 & 6 & 9 & 8 & 13.9\% & 9.5\%--20.0\% \\
Signer repository & 1 & 3 & 4 & 6 & 7.9\% & 4.7\%--13.0\% \\\midrule
Packages & 28 & 42 & 60 & 63 & \multicolumn{2}{c@{}}{165} \\
\bottomrule
\end{tabular*}
\par\smallskip\parbox{\columnwidth}{\scriptsize\raggedright Packages with at least one alert (timestamp predecessor, adjacent memory). Denominator (last row): the packages with at least two keyless-signed releases, since a pin needs one release to learn the signer and a second to compare; Wilson intervals. S1--S4: samples~1--4. Signer repository: the repository of the workflow that the SAN names. Pooled shares by registry, in row order: npm 67.1\%, 48.1\%, 20.3\% and 10.1\% of 79 packages; PyPI 83.7\%, 14.0\%, 8.1\% and 5.8\% of 86 packages. A pin on the certificate's source-repository field alerts in 12 of the 165 pooled packages.}
\end{table}

%% file: refs.bib
@inproceedings{halder2024metadata,
  author       = {Sajal Halder and Michael Bewong and Arash Mahboubi and Yinhao Jiang and Md. Rafiqul Islam and Md Zahidul Islam and Ryan H. L. Ip and Muhammad Ejaz Ahmed and Gowri Sankar Ramachandran and Muhammad Ali Babar},
  title        = {Malicious Package Detection using Metadata Information},
  booktitle    = {Proceedings of the ACM Web Conference 2024 (WWW '24)},
  year         = {2024},
  pages        = {1779--1789},
  publisher    = {ACM},
  doi          = {10.1145/3589334.3645543}
}

@inproceedings{zimmermann2019,
  author       = {Markus Zimmermann and Cristian-Alexandru Staicu and Cam Tenny and Michael Pradel},
  title        = {Small World with High Risks: A Study of Security Threats in the {npm} Ecosystem},
  booktitle    = {Proceedings of the 28th USENIX Security Symposium (USENIX Security '19)},
  year         = {2019},
  pages        = {995--1010},
  publisher    = {USENIX Association}
}

@inproceedings{torresarias2019intoto,
  author       = {Santiago Torres-Arias and Hammad Afzali and Trishank Karthik Kuppusamy and Reza Curtmola and Justin Cappos},
  title        = {{in-toto}: Providing farm-to-table guarantees for bits and bytes},
  booktitle    = {Proceedings of the 28th USENIX Security Symposium (USENIX Security '19)},
  year         = {2019},
  pages        = {1393--1410},
  publisher    = {USENIX Association}
}

@inproceedings{kalu2025industrysigning,
  author       = {Kelechi G. Kalu and Tanmay Singla and Chinenye Okafor and Santiago Torres-Arias and James C. Davis},
  title        = {An Industry Interview Study of Software Signing for Supply Chain Security},
  booktitle    = {Proceedings of the 34th USENIX Security Symposium (USENIX Security '25)},
  year         = {2025},
  pages        = {81--100},
  publisher    = {USENIX Association}
}

@inproceedings{kalu2026usability,
  author       = {Kelechi G. Kalu and Sofia Okorafor and Tanmay Singla and Sophie Chen and Santiago Torres-Arias and James C. Davis},
  title        = {Why {Johnny} Adopts Identity-Based Software Signing: A Usability Case Study of {Sigstore}},
  booktitle    = {Proceedings of the 35th USENIX Security Symposium (USENIX Security '26)},
  year         = {2026},
  pages        = {5433--5452},
  publisher    = {USENIX Association}
}

@inproceedings{kuppusamy2016diplomat,
  author       = {Trishank Karthik Kuppusamy and Santiago Torres-Arias and Vladimir Diaz and Justin Cappos},
  title        = {{Diplomat}: Using Delegations to Protect Community Repositories},
  booktitle    = {Proceedings of the 13th USENIX Symposium on Networked Systems Design and Implementation (NSDI '16)},
  year         = {2016},
  pages        = {567--581},
  publisher    = {USENIX Association}
}

@inproceedings{newman2022sigstore,
  author       = {Zachary Newman and John Speed Meyers and Santiago Torres-Arias},
  title        = {{Sigstore}: Software Signing for Everybody},
  booktitle    = {Proceedings of the 2022 ACM SIGSAC Conference on Computer and Communications Security (CCS '22)},
  year         = {2022},
  pages        = {2353--2367},
  publisher    = {ACM},
  doi          = {10.1145/3548606.3560596}
}

@inproceedings{merrill2023speranza,
  author       = {Kelsey Merrill and Zachary Newman and Santiago Torres-Arias and Karen R. Sollins},
  title        = {{Speranza}: Usable, Privacy-friendly Software Signing},
  booktitle    = {Proceedings of the 2023 ACM SIGSAC Conference on Computer and Communications Security (CCS '23)},
  year         = {2023},
  pages        = {3388--3402},
  publisher    = {ACM},
  doi          = {10.1145/3576915.3623200}
}

@inproceedings{cappos2008mirror,
  author       = {Justin Cappos and Justin Samuel and Scott Baker and John H. Hartman},
  title        = {A Look in the Mirror: Attacks on Package Managers},
  booktitle    = {Proceedings of the 15th ACM Conference on Computer and Communications Security (CCS '08)},
  year         = {2008},
  pages        = {565--574},
  publisher    = {ACM},
  doi          = {10.1145/1455770.1455841}
}

@inproceedings{gu2023registries,
  author       = {Yacong Gu and Lingyun Ying and Yingyuan Pu and Xiao Hu and Huajun Chai and Ruimin Wang and Xing Gao and Haixin Duan},
  title        = {Investigating Package Related Security Threats in Software Registries},
  booktitle    = {2023 IEEE Symposium on Security and Privacy (SP)},
  year         = {2023},
  pages        = {1578--1595},
  publisher    = {IEEE},
  doi          = {10.1109/SP46215.2023.10179332}
}

@inproceedings{schorlemmer2024registrysigning,
  author       = {Taylor R. Schorlemmer and Kelechi G. Kalu and Luke Chigges and Kyung Myung Ko and Eman Abu Ishgair and Saurabh Bagchi and Santiago Torres-Arias and James C. Davis},
  title        = {Signing in Four Public Software Package Registries: Quantity, Quality, and Influencing Factors},
  booktitle    = {2024 IEEE Symposium on Security and Privacy (SP)},
  year         = {2024},
  pages        = {1160--1178},
  publisher    = {IEEE},
  doi          = {10.1109/SP54263.2024.00215}
}

@inproceedings{ladisa2023sok,
  author       = {Piergiorgio Ladisa and Henrik Plate and Matias Martinez and Olivier Barais},
  title        = {{SoK}: Taxonomy of Attacks on Open-Source Software Supply Chains},
  booktitle    = {2023 IEEE Symposium on Security and Privacy (SP)},
  year         = {2023},
  pages        = {1509--1526},
  publisher    = {IEEE},
  doi          = {10.1109/SP46215.2023.10179304}
}

@inproceedings{duan2021measuring,
  author       = {Ruian Duan and Omar Alrawi and Ranjita Pai Kasturi and Ryan Elder and Brendan Saltaformaggio and Wenke Lee},
  title        = {Towards Measuring Supply Chain Attacks on Package Managers for Interpreted Languages},
  booktitle    = {Proceedings 2021 Network and Distributed System Security Symposium},
  year         = {2021},
  publisher    = {Internet Society},
  doi          = {10.14722/ndss.2021.23055}
}

@inproceedings{kranch2015pinning,
  author       = {Michael Kranch and Joseph Bonneau},
  title        = {Upgrading {HTTPS} in Mid-Air: An Empirical Study of Strict Transport Security and Key Pinning},
  booktitle    = {Proceedings 2015 Network and Distributed System Security Symposium},
  year         = {2015},
  publisher    = {Internet Society},
  doi          = {10.14722/ndss.2015.23162}
}

@inproceedings{ohm2020backstabbers,
  author       = {Marc Ohm and Henrik Plate and Arnold Sykosch and Michael Meier},
  title        = {Backstabber's Knife Collection: A Review of Open Source Software Supply Chain Attacks},
  booktitle    = {Detection of Intrusions and Malware, and Vulnerability Assessment},
  year         = {2020},
  pages        = {23--43},
  publisher    = {Springer International Publishing},
  doi          = {10.1007/978-3-030-52683-2_2}
}

@inproceedings{sofaer2024rogueone,
  author       = {Raphael J. Sofaer and Yaniv David and Mingqing Kang and Jianjia Yu and Yinzhi Cao and Junfeng Yang and Jason Nieh},
  title        = {{RogueOne}: Detecting Rogue Updates via Differential Data-flow Analysis Using Trust Domains},
  booktitle    = {Proceedings of the IEEE/ACM 46th International Conference on Software Engineering (ICSE '24)},
  year         = {2024},
  pages        = {1--13},
  publisher    = {ACM},
  doi          = {10.1145/3597503.3639199}
}

@inproceedings{zahan2022weaklinks,
  author       = {Nusrat Zahan and Thomas Zimmermann and Patrice Godefroid and Brendan Murphy and Chandra Maddila and Laurie Williams},
  title        = {What Are Weak Links in the {npm} Supply Chain?},
  booktitle    = {Proceedings of the 44th International Conference on Software Engineering: Software Engineering in Practice (ICSE-SEIP '22)},
  year         = {2022},
  pages        = {331--340},
  publisher    = {ACM},
  doi          = {10.1145/3510457.3513044}
}

@inproceedings{guo2023malicious,
  author       = {Wenbo Guo and Zhengzi Xu and Chengwei Liu and Cheng Huang and Yong Fang and Yang Liu},
  title        = {An Empirical Study of Malicious Code In {PyPI} Ecosystem},
  booktitle    = {2023 38th IEEE/ACM International Conference on Automated Software Engineering (ASE)},
  year         = {2023},
  pages        = {166--177},
  publisher    = {IEEE},
  doi          = {10.1109/ASE56229.2023.00135}
}

@inproceedings{amann2017diginotar,
  author       = {Johanna Amann and Oliver Gasser and Quirin Scheitle and Lexi Brent and Georg Carle and Ralph Holz},
  title        = {Mission Accomplished? {HTTPS} Security after {DigiNotar}},
  booktitle    = {Proceedings of the 2017 Internet Measurement Conference (IMC '17)},
  year         = {2017},
  pages        = {325--340},
  publisher    = {ACM},
  doi          = {10.1145/3131365.3131401}
}

@inproceedings{scheitle2018ct,
  author       = {Quirin Scheitle and Oliver Gasser and Theodor Nolte and Johanna Amann and Lexi Brent and Georg Carle and Ralph Holz and Thomas C. Schmidt and Matthias W{\"a}hlisch},
  title        = {The Rise of Certificate Transparency and Its Implications on the Internet Ecosystem},
  booktitle    = {Proceedings of the Internet Measurement Conference 2018 (IMC '18)},
  year         = {2018},
  pages        = {343--349},
  publisher    = {ACM},
  doi          = {10.1145/3278532.3278562}
}

@article{scheitle2018caa,
  author       = {Quirin Scheitle and Taejoong Chung and Jens Hiller and Oliver Gasser and Johannes Naab and Roland van Rijswijk-Deij and Oliver Hohlfeld and Ralph Holz and Dave Choffnes and Alan Mislove and Georg Carle},
  title        = {A First Look at Certification Authority Authorization ({CAA})},
  journal      = {ACM SIGCOMM Computer Communication Review},
  year         = {2018},
  volume       = {48},
  number       = {2},
  pages        = {10--23},
  doi          = {10.1145/3213232.3213235}
}

@inproceedings{guo2026npmdetectors,
  author       = {Wenbo Guo and Zhongwen Chen and Zhengzi Xu and Chengwei Liu and Ming Kang and Shiwen Song and Chengyue Liu and Yijia Xu and Weisong Sun and Yang Liu},
  title        = {How Effective Are {NPM} Malicious Package Detectors? A Large-Scale Empirical Study},
  booktitle    = {Proceedings of the 41st IEEE/ACM International Conference on Automated Software Engineering},
  year         = {2026},
  pages        = {1651--1663},
  publisher    = {ACM},
  doi          = {10.1145/3832783.3837427}
}

@inproceedings{okafor2024diverify,
  author       = {Chinenye Okafor and James C. Davis and Santiago Torres-Arias},
  title        = {Context-Aware Trust Verification for Identity-Based Software Signing},
  booktitle    = {Proceedings of the 41st IEEE/ACM International Conference on Automated Software Engineering},
  year         = {2026},
  pages        = {29--41},
  publisher    = {ACM},
  doi          = {10.1145/3832783.3834333}
}

@misc{tanaka2026cooldown,
  author       = {Hidetake Tanaka and Rikuto Tsuchida and Kazumasa Shimari and Raula Gaikovina Kula and Kenichi Matsumoto},
  title        = {An Exploratory Study of {Dependabot} Cooldown Adoption in Open-Source {GitHub} Projects},
  year         = {2026},
  howpublished = {arXiv preprint arXiv:2609.16605},
  url          = {https://arxiv.org/abs/2609.16605}
}

@misc{rfc6797,
  author       = {Jeff Hodges and Collin Jackson and Adam Barth},
  title        = {{HTTP} Strict Transport Security ({HSTS})},
  year         = {2012},
  howpublished = {RFC 6797},
  doi          = {10.17487/RFC6797}
}

@misc{rfc7469,
  author       = {Chris Evans and Chris Palmer and Ryan Sleevi},
  title        = {Public Key Pinning Extension for {HTTP}},
  year         = {2015},
  howpublished = {RFC 7469},
  doi          = {10.17487/RFC7469}
}

@misc{rfc8659,
  author       = {Phillip Hallam-Baker and Rob Stradling and Jacob Hoffman-Andrews},
  title        = {{DNS} Certification Authority Authorization ({CAA}) Resource Record},
  year         = {2019},
  howpublished = {RFC 8659},
  doi          = {10.17487/RFC8659}
}

@misc{pep458,
  author       = {Trishank Karthik Kuppusamy and Vladimir Diaz and Marina Moore and Lukas Puehringer and Joshua Lock and Lois Anne DeLong and Justin Cappos},
  title        = {{PEP} 458 -- Secure {PyPI} downloads with signed repository metadata},
  year         = {2013},
  url          = {https://peps.python.org/pep-0458/},
  note         = {Accessed 9 October 2026}
}

@misc{pep480,
  author       = {Trishank Karthik Kuppusamy and Vladimir Diaz and Justin Cappos and Marina Moore},
  title        = {{PEP} 480 -- Surviving a Compromise of {PyPI}: End-to-end signing of packages},
  year         = {2014},
  url          = {https://peps.python.org/pep-0480/},
  note         = {Accessed 9 October 2026}
}

@misc{pep740,
  author       = {William Woodruff and Facundo Tuesca and Dustin Ingram},
  title        = {{PEP} 740 -- Index support for digital attestations},
  year         = {2024},
  url          = {https://peps.python.org/pep-0740/},
  note         = {Accessed 9 October 2026}
}

@misc{pep751,
  author       = {Brett Cannon},
  title        = {{PEP} 751 -- A file format to record {Python} dependencies for installation reproducibility},
  year         = {2024},
  url          = {https://peps.python.org/pep-0751/},
  note         = {Accessed 4 October 2026}
}

@misc{slsa2025v12,
  author       = {{SLSA}},
  title        = {{SLSA} specification},
  year         = {2025},
  howpublished = {Version 1.2},
  url          = {https://slsa.dev/spec/v1.2/},
  note         = {Accessed 9 October 2026}
}

@misc{tufspec2026,
  author       = {{The Update Framework}},
  title        = {The Update Framework Specification},
  year         = {2026},
  howpublished = {Version 1.0.36},
  url          = {https://theupdateframework.github.io/specification/v1.0.36/},
  note         = {Accessed 9 October 2026}
}

@misc{openssf2024principles,
  author       = {Jack Cable and Zach Steindler},
  title        = {Principles for Package Repository Security},
  year         = {2024},
  howpublished = {OpenSSF Securing Software Repositories Working Group},
  url          = {https://repos.openssf.org/principles-for-package-repository-security},
  note         = {Accessed 9 October 2026}
}

@misc{cabforum2017ballot187,
  author       = {{CA/Browser Forum}},
  title        = {Ballot 187 -- Make {CAA} Checking Mandatory},
  year         = {2017},
  url          = {https://cabforum.org/2017/03/08/ballot-187-make-caa-checking-mandatory/},
  note         = {Accessed 9 October 2026}
}

@misc{chrome2026ctpolicy,
  author       = {{Google Chrome}},
  title        = {Chrome Certificate Transparency Policy},
  year         = {n.d.},
  url          = {https://googlechrome.github.io/CertificateTransparency/ct_policy.html},
  note         = {Accessed 9 October 2026}
}

@misc{npm_trusted_publishers,
  author       = {{npm}},
  title        = {Trusted publishing for {npm} packages},
  year         = {n.d.},
  url          = {https://docs.npmjs.com/trusted-publishers/},
  note         = {Accessed 9 October 2026}
}

@misc{npm_provenance,
  author       = {{npm}},
  title        = {Generating provenance statements},
  year         = {n.d.},
  url          = {https://docs.npmjs.com/generating-provenance-statements/},
  note         = {Accessed 9 October 2026}
}

@misc{npm_registry_signatures,
  author       = {{npm}},
  title        = {About {ECDSA} registry signatures},
  year         = {n.d.},
  url          = {https://docs.npmjs.com/about-registry-signatures/},
  note         = {Accessed 9 October 2026}
}

@misc{npm_staged_publishing,
  author       = {{npm}},
  title        = {Staged publishing for {npm} packages},
  year         = {n.d.},
  url          = {https://docs.npmjs.com/staged-publishing/},
  note         = {Accessed 9 October 2026}
}

@misc{pypi_trusted_publishers,
  author       = {{PyPI}},
  title        = {Publishing to {PyPI} with a Trusted Publisher},
  year         = {n.d.},
  url          = {https://docs.pypi.org/trusted-publishers/},
  note         = {Accessed 9 October 2026}
}

@misc{crates2026trusted,
  author       = {Tobias Bieniek},
  title        = {crates.io: development update},
  year         = {2025},
  howpublished = {Rust Blog},
  url          = {https://blog.rust-lang.org/2025/07/11/crates-io-development-update-2025-07/},
  note         = {Accessed 9 October 2026}
}

@misc{rubygems2026api,
  author       = {{RubyGems}},
  title        = {{RubyGems.org} {API}},
  year         = {n.d.},
  url          = {https://guides.rubygems.org/rubygems-org-api/},
  note         = {Accessed 9 October 2026}
}

@misc{sonatype2026namespace,
  author       = {{Sonatype}},
  title        = {Register a Namespace},
  year         = {n.d.},
  url          = {https://central.sonatype.org/register/namespace/},
  note         = {Accessed 9 October 2026}
}

@misc{sonatype2026requirements,
  author       = {{Sonatype}},
  title        = {Requirements},
  year         = {n.d.},
  url          = {https://central.sonatype.org/publish/requirements/},
  note         = {Accessed 9 October 2026}
}

@misc{sonatype2025sigstoreportal,
  author       = {{Sonatype}},
  title        = {{Sigstore} Signature Validation Is Now Available Via The Central Publisher Portal},
  year         = {2025},
  url          = {https://central.sonatype.org/news/20250128_sigstore_signature_validation_via_portal/},
  note         = {Accessed 9 October 2026}
}

@misc{apache_release_distribution,
  author       = {{The Apache Software Foundation}},
  title        = {Release Distribution Policy},
  year         = {n.d.},
  url          = {https://infra.apache.org/release-distribution.html},
  note         = {Accessed 9 October 2026}
}

@misc{fulcio_oid,
  author       = {{Sigstore}},
  title        = {{Sigstore} {OID} information},
  year         = {n.d.},
  url          = {https://github.com/sigstore/fulcio/blob/main/docs/oid-info.md},
  note         = {Accessed 9 October 2026}
}

@misc{cosign_verify,
  author       = {{Sigstore}},
  title        = {cosign verify},
  year         = {n.d.},
  url          = {https://github.com/sigstore/cosign/blob/main/doc/cosign_verify.md},
  note         = {Accessed 22 October 2026}
}

@misc{gh_attestation_verify,
  author       = {{GitHub}},
  title        = {gh attestation verify},
  year         = {n.d.},
  url          = {https://cli.github.com/manual/gh_attestation_verify},
  note         = {Accessed 9 October 2026}
}

@misc{gradle_verification,
  author       = {{Gradle}},
  title        = {Verifying Dependencies},
  year         = {n.d.},
  url          = {https://docs.gradle.org/current/userguide/dependency_verification.html},
  note         = {Accessed 9 October 2026}
}

@misc{socket_alerts,
  author       = {{Socket}},
  title        = {Alerts},
  year         = {n.d.},
  url          = {https://socket.dev/alerts},
  note         = {Accessed 9 October 2026}
}

@misc{pnpm_trust_policy,
  author       = {{pnpm}},
  title        = {Dependency Resolution Settings},
  year         = {n.d.},
  url          = {https://pnpm.io/settings/dependency-resolution},
  note         = {Accessed 9 October 2026}
}

@misc{pnpm_prerelease_fix,
  author       = {{pnpm}},
  title        = {{pnpm} 10.24},
  year         = {2025},
  url          = {https://pnpm.io/blog/releases/10.24},
  note         = {Accessed 9 October 2026}
}

@misc{pnpm_releases,
  author       = {{pnpm}},
  title        = {Releases: pnpm/pnpm},
  year         = {n.d.},
  url          = {https://github.com/pnpm/pnpm/releases},
  note         = {Accessed 9 October 2026}
}

@misc{pnpm_issue_10202,
  author       = {Hyunbin Seo},
  title        = {{no-downgrade} should check the same major version},
  year         = {2025},
  url          = {https://github.com/pnpm/pnpm/issues/10202},
  note         = {Accessed 9 October 2026}
}

@misc{pnpm_pr_10359,
  author       = {{btea}},
  title        = {feat: add {trustPolicyIgnoreAfter}},
  year         = {2025},
  url          = {https://github.com/pnpm/pnpm/pull/10359},
  note         = {Accessed 9 October 2026}
}

@misc{pnpm_issue_11887,
  author       = {Kevin Deng},
  title        = {no-downgrade false positive with staged publishing},
  year         = {2026},
  url          = {https://github.com/pnpm/pnpm/issues/11887},
  note         = {Accessed 9 October 2026}
}

@misc{githubchangelog2025npmtrusted,
  author       = {{GitHub}},
  title        = {{npm} trusted publishing with {OIDC} is generally available},
  year         = {2025},
  url          = {https://github.blog/changelog/2025-07-31-npm-trusted-publishing-with-oidc-is-generally-available/},
  note         = {Accessed 9 October 2026}
}

@misc{crates2026trustpubonly,
  author       = {Tobias Bieniek},
  title        = {crates.io: development update},
  year         = {2026},
  howpublished = {Rust Blog},
  url          = {https://blog.rust-lang.org/2026/01/21/crates-io-development-update/},
  note         = {Accessed 9 October 2026}
}

@misc{palmer2017hpkp,
  author       = {Chris Palmer},
  title        = {Intent To Deprecate And Remove: Public Key Pinning},
  year         = {2017},
  howpublished = {blink-dev mailing list},
  url          = {https://groups.google.com/a/chromium.org/g/blink-dev/c/he9tr7p3rZ8},
  note         = {Accessed 9 October 2026}
}

@misc{chromestatus2018hpkp,
  author       = {{Chrome Platform Status}},
  title        = {Remove {HTTP}-Based Public Key Pinning},
  year         = {2018},
  url          = {https://chromestatus.com/feature/5903385005916160},
  note         = {Accessed 9 October 2026}
}

@misc{microsoft2025shaihulud,
  author       = {{Microsoft Defender Security Research Team} and Veronica Sun},
  title        = {{Shai-Hulud} 2.0: Guidance for detecting, investigating, and defending against the supply chain attack},
  year         = {2025},
  url          = {https://www.microsoft.com/en-us/security/blog/2025/12/09/shai-hulud-2-0-guidance-for-detecting-investigating-and-defending-against-the-supply-chain-attack/},
  note         = {Accessed 22 October 2026}
}

@misc{posthog2025postmortem,
  author       = {Oliver Browne},
  title        = {Post-mortem of {Shai-Hulud} attack on {November} 24th, 2025},
  year         = {2025},
  howpublished = {PostHog Blog},
  url          = {https://posthog.com/blog/nov-24-shai-hulud-attack-post-mortem},
  note         = {Accessed 9 October 2026}
}

@misc{axios2026postmortem,
  author       = {Jason Saayman},
  title        = {Post Mortem: {axios} {npm} supply chain compromise},
  year         = {2026},
  url          = {https://github.com/axios/axios/issues/10636},
  note         = {Accessed 9 October 2026}
}

@misc{pypi2026litellm_telnyx,
  author       = {Seth Larson and Mike Fiedler},
  title        = {Incident Report: {LiteLLM}/{Telnyx} supply-chain attacks, with guidance},
  year         = {2026},
  howpublished = {The Python Package Index Blog},
  url          = {https://blog.pypi.org/posts/2026-04-02-incident-report-litellm-telnyx-supply-chain-attack/},
  note         = {Accessed 9 October 2026}
}
